\RequirePackage[mathlines]{lineno} 
\documentclass[prd,twocolumn,showpacs,amsmath,amssymb]{revtex4-1}
\usepackage{overpic,graphicx}
\usepackage{dcolumn}
\usepackage{bm}
\usepackage{rotating}
\usepackage{subfigure}
\usepackage{color}
\usepackage{dcolumn}
\usepackage{siunitx}
\usepackage{graphicx,stackengine}
\usepackage{lipsum} 
\usepackage{appendix}
\usepackage{lineno}

\renewcommand{\thetable}{\arabic{table}}

\def \c0  {\chi_{c0}}
\def \c1  {\chi_{c1}}
\def \c2  {\chi_{c2}}

\def \ee   {e^+e^-}
\def \uu   {\mu^+\mu^-}
\def \gevc {\mbox{GeV/$c$}}
\def \gevcc{\mbox{GeV/$c^2$}}
\def \mev  {\mbox{MeV}}

\def \mevcc{\mbox{MeV/$c^2$}}

\def \jpsi {J/\psi}

\begin{document}

\title{\bf \boldmath Measurement of $\ee\to\gamma \chi_{c0,c1,c2}$ cross sections at center-of-mass energies between 3.77 and 4.60~GeV}

\author{
M.~Ablikim$^{1}$, M.~N.~Achasov$^{10,b}$, P.~Adlarson$^{67}$, S. ~Ahmed$^{15}$, M.~Albrecht$^{4}$, R.~Aliberti$^{28}$, A.~Amoroso$^{66A,66C}$, M.~R.~An$^{32}$, Q.~An$^{63,49}$, X.~H.~Bai$^{57}$
% ...
\\
\vspace{0.2cm}
(BESIII Collaboration)\\
\vspace{0.2cm} {\it
	$^{1}$ Institute of High Energy Physics, Beijing 100049, People's Republic of China\\
% ...
$^{4}$ Bochum Ruhr-University, D-44780 Bochum, Germany\\
% ...
$^{10}$ G.I. Budker Institute of Nuclear Physics SB RAS (BINP), Novosibirsk 630090, Russia\\
% ...
$^{15}$ Helmholtz Institute Mainz, Staudinger Weg 18, D-55099 Mainz, Germany\\
% ...
$^{28}$ Johannes Gutenberg University of Mainz, Johann-Joachim-Becher-Weg 45, D-55099 Mainz, Germany\\
% ...
$^{32}$ Liaoning Normal University, Dalian 116029, People's Republic of China\\
% ...
$^{49}$ State Key Laboratory of Particle Detection and Electronics, Beijing 100049, Hefei 230026, People's Republic of China\\
% ...
$^{57}$ University of Jinan, Jinan 250022, People's Republic of China\\
% ...
$^{63}$ University of Science and Technology of China, Hefei 230026, People's Republic of China\\
% ...
$^{66}$ University of Turin and INFN, (A)University of Turin, I-10125, Turin, Italy; (B)University of Eastern Piedmont, I-15121, Alessandria, Italy; (C)INFN, I-10125, Turin, Italy\\
	$^{67}$ Uppsala University, Box 516, SE-75120 Uppsala, Sweden\\
% ...
$^{b}$ Also at the Novosibirsk State University, Novosibirsk, 630090, Russia\\
% ...
	}
	\vspace{0.4cm}
}


\begin{abstract}

The $\ee\to\gamma\chi_{cJ}$ ($J=0,1,2$) processes are studied  at center-of-mass energies ranging from 3.773 to 4.600~GeV, using a total integrated luminosity of 19.3~fb$^{-1}$ 
$\ee$ annihilation data accumulated with the BESIII detector at BEPCII. We observe for the first time $\ee\to\gamma \chi_{c1,c2}$ signals at $\sqrt{s}=$ 4.180~GeV with 
statistical significances of 7.6$\sigma$ and 6.0$\sigma$, respectively. 
The production cross section of $\ee\to\gamma \chi_{c1,c2}$ at each center-of-mass energy is also measured.
We find that the line shape of the $\ee\to\gamma \chi_{c1}$ cross section can be described with conventional charmonium states 
$\psi(3686)$, $\psi(3770)$, $\psi(4040)$, $\psi(4160)$. Compared with this, for the $\ee\to\gamma \chi_{c2}$ channel, one more additional 
resonance is added to describe the cross section line shape. Its mass and width are measured to be 
$M=4371.7\pm7.5\pm1.8$~$\mevcc$ and $\Gamma^{tot}=51.1\pm17.6\pm1.9$~$\mev$, where the first uncertainties are statistical and 
the second systematic. The significance of this resonance is estimated to be 5.8$\sigma$, and its parameters agree with the $Y(4360)$ resonance previously reported in $\ee\to\pi^+\pi^-\psi(3686)$, and the $Y(4390)$ in $\ee\to\pi^+\pi^-h_c$ within uncertainties.
No significant signal for the $\ee\to\gamma \chi_{c0}$ process is observed, and the upper limits of Born cross sections
$\sigma_{B}(\ee\to\gamma \chi_{c0})$ at 90\% confidence level are reported. 


\end{abstract}


\oddsidemargin  -0.2cm
\evensidemargin -0.2cm
\maketitle
\section{INTRODUCTION}
In the past decades, many charmoniumlike states were observed experimentally, such as the $X(3872)$, $Y(4260)$, and $Z_c(3900)$~\cite{Olsen:2017bmm}.
Among them, the vector $Y$-states should have quantum numbers $J^{PC} = 1^{--}$, as they are produced in $\ee$ annihilation process. Considering the $Y(4260)$~\cite{Aubert:2005rm,He:2006kg,Yuan:2007sj,Yuan:2007sj,Liu:2013dau}, 
$Y(4360)$, and $Y(4660)$ states~\cite{Aubert:2006ge,Wang:2007ea,Lees:2012pv,Wang:2014hta}, together with the conventional charmonium states $\psi(4040)$, $\psi(4160)$, and $\psi(4415)$, there are at least six vector states
between 4.0 and 4.7~GeV. However, the potential model only predicts five vector charmonium states in this mass region~\cite{potential}. In addition, unlike the known 
$1^{--}$ conventional charmonium states that decay predominantly into open-charm final states [$D^{(*)}\overline{D}^{(*)}$], the $Y$-states show strong coupling to 
hidden-charm final states. These unusual behaviors indicate that the $Y$-states might be non-conventional quarkonium states. To better understand the nature of these states and also one gets better insights in the relevant degrees of freedom that play a role in these systems that are governed by the strong interaction, it is important to further investigate these states experimentally.

The radiative transition rates between charmonium states have been predicted theoretically from potential models~\cite{Barnes:2005pb}. The partial widths of electric dipole (E1) transitions between $\psi(4040)/\psi(4160)/\psi(4415)$ and $\chi_{cJ}$ states ($J=0,1,2$) are in the range 0 $\sim$ 35~keV. Quoting the full width of $\psi(4040)$, $\psi(4160)$, and $\psi(4415)$ to be 80, 70, and 62~MeV~\cite{pdg}, respectively, and the expected branching fractions are at the level of  $10^{-7}$ - $10^{-4}$. By studying the radiative transitions between vector $Y$-states and $\chi_{cJ}$ ($J=0,1,2$), we can compare the decay of $Y$-states with conventional charmonium states, and thus help to understand the nature of $Y$ states~\cite{Ma:2014ofa,Chao:2013cca}.

Experimentally, the $\ee\to\gamma\chi_{cJ}$ ($J=1,2$) processes above 4~GeV have been studied before by BESIII~\cite{Ablikim:2014hwn}, CLEO~\cite{CLEOchicj}, and Belle experiments~\cite{Bellechicj}. Due to the limited statistics, no obvious signal has been observed between 4 - 5~GeV. 
The BESIII has collected the world's largest dataset from 4.0 to 4.6~GeV, and it is thus highly motivated to search for these decay modes.

In this paper, we report the study of the $\ee\to\gamma \chi_{cJ} (J=0,1,2)$ processes at $\ee$ center-of-mass (c.m.) energies between $\sqrt{s}=$ 4.008 - 4.6~GeV, using data samples corresponding to an integrated luminosity of 16.0~fb$^{-1}$ accumulated with the BESIII detector at the BEPCII collider. To better estimate the contributions from $\psi(3686)$ and $\psi(3770)$, the datasets with integrated luminosity of 3.3~fb$^{-1}$ between $\sqrt{s}$ of 3.773 and 4.008~GeV for $\ee\to\gamma \chi_{c1,c2}$ channels are also analyzed. The datasets together with the corresponding c.m. energies are summarized in Table~\ref{tab:cross-db_c1} in the Appendix. Compared with the previous BESIII measurement~\cite{Ablikim:2014hwn}, the new dataset covers an extended c.m. energies with about one order of magnitude higher luminosity, and also both $\jpsi\to e^{+}e^{-}/\mu^+\mu^-$ events (only $\mu^+\mu^-$ used in previous work) are studied.
The integrated luminosities are measured with Bhabha events ($\ee\to(\gamma)\ee$) with an uncertainty of 1$\%$~\cite{Ablikim:2015nan}. The c.m. energy of each dataset is measured using dimuon events ($\ee\to(\gamma)\uu$), with an uncertainty of $\pm$ 0.8~MeV~\cite{Ablikim:2015zaa}. 

\section{BESIII DETECTOR AND MC SIMULATION}
The BESIII detector~\cite{Ablikim:2009aa} records symmetric $e^+e^-$ collisions 
provided by the BEPCII storage ring~\cite{Yu:IPAC2016-TUYA01}, with a designed peak luminosity of $1\times10^{33}$~cm$^{-2}$s$^{-1}$ at c.m. energy of 3.77~GeV.
BESIII has collected large data samples between 2.0 and 4.6~GeV~\cite{Ablikim:2019hff}. The cylindrical core of the BESIII detector covers 93\% of the full solid angle and consists of a helium-based multilayer drift chamber~(MDC), a plastic scintillator time-of-flight
system~(TOF), and a CsI(Tl) electromagnetic calorimeter~(EMC), which are all enclosed in a superconducting solenoidal magnet providing a 1.0~T magnetic field. The solenoid is supported by an
octagonal flux-return yoke with resistive plate counter muon
identification modules interleaved with steel. 
The charged-particle momentum resolution at $1~{\rm GeV}/c$ is
$0.5\%$, and the d$E/$d$x$ resolution is $6\%$ for electrons
from Bhabha scattering. The EMC measures photon energies with a
resolution of $2.5\%$ ($5\%$) at $1$~GeV in the barrel (end cap)
region~\cite{Ablikim:2009aa}. The time resolution in the TOF barrel region is 68~ps, while
that in the end cap region is 110~ps. The end cap TOF
	system was upgraded in 2015 using multi-gap resistive plate chamber
	technology, providing a time resolution of
	60~ps~\cite{etof}.

Simulated Monte Carlo (MC) samples produced with {\sc geant4}-based~\cite{geant4} software, which includes the geometrical description of the BESIII detector and the detector response, are used to determine the detection efficiency, and to estimate physical background. The signal MC $\ee\to\gamma\chi_{c0,c1,c2}$ events are generated assuming a pure E1 transition.  The simulation models the beam energy spread and initial-state-radiation (ISR) in $e^+e^-$ annihilation using the generator {\sc	kkmc}~\cite{ref:kkmc}. The maximum ISR photon energy is set to the energy corresponding to the $\gamma\chi_{c0,c1,c2}$ production threshold. The final-state-radiation (FSR) from charged final state particles is modelled with {\sc 	photos}~\cite{photos}. Possible background contributions are investigated with the inclusive MC samples, which consist of open-charm processes, the ISR production of lower mass vector charmonium(-like) states, and the continuum processes. The known decay modes of charmed hadrons are modelled with {\sc 	evtgen}~\cite{ref:evtgen}, with known  branching fractions taken from the Particle Data Group (PDG)~\cite{pdg}, and the remaining unknown decays with {\sc lundcharm}~\cite{ref:lundcharm}. 

\section{$\ee\to\gamma\chi_{c1,c2}$}
\subsection{Event selection}
 The final state particles for $\ee\to\gamma\chi_{c1,c2}$ are $\gamma\gamma\ell^{+}\ell^{-}$, where the $\chi_{c1,c2}$ are reconstructed with $\gamma\jpsi$, and the $\jpsi$ is reconstructed with $\ell^{+}\ell^{-}$ ($\ell=$ e or $\mu$).  Events with two charged tracks with zero net charge and at least two photons are selected. Each charged track is required to originate from the interaction point, within $\pm$1~cm in the plane perpendicular to the beams and 10~cm along the beam direction. The $|\!\cos\theta|$ of each charged track is required to be less than 0.93, where $\theta$ is the polar angle of each track. Photons are required to have a deposited energy larger than 25~$\mev$ in the barrel EMC region ($|\!\cos\theta|<0.8$) and larger than 50~$\mev$ in the end cap region ($0.86<|\!\cos\theta|<0.92$). The EMC time for a photon is required to be within 700~ns of the event start time to suppress the electronic noise and energy deposition unrelated to the physical events. Each charged track should have a momentum larger than 1~$\gevc$. For leptons, we use the energy deposited in the EMC to separate electrons from muons. Charged tracks with the energy deposited in the EMC larger than 1 GeV are identified as electrons, and charged tracks with the energy deposited less than 0.4~GeV are identified as muons. For photons, the two most energetic photons are regarded as the candidates for signal events. Through the paper, we denote the photon with higher energy as $\gamma_{H}$, and the other as $\gamma_{L}$. 

 To improve the mass resolution and to suppress backgrounds, a four-constraint (4C) kinematic fit is performed under the $\gamma\gamma\ell^{+}\ell^{-}$ hypothesis, which constrains the total four momentum of the final measured particles to the initial four-momentum of the colliding beams. The $\chi^{2}$ of the kinematic fit is required to be less than 40. 	 

To suppress the radiative Bhabha events ($\ee\to \gamma\ee$) in the $\jpsi\to\ee$ mode, the cosine of the opening angle between the electron and the nearest photon ($\!\cos\theta_{e\gamma})$ is required to be less than 0.86. Since the photon from radiative Bhabha process is always close to the beam direction, the cosine of the polar angle of the selected photons are required to satisfy $|\!\cos\theta_{\gamma_{L/H}}|<0.8$.
In both $\ee$ and $\uu$ modes, the background from $\ee\to\eta\jpsi$ with $\eta\to\gamma\gamma$ is rejected by requiring $|M(\gamma_{H}\gamma_{L})-m(\eta)|>0.03$~$\gevcc$. Furthermore, the $\pi^{+}\pi^{-}\pi^{0}$ background is rejected by requiring $|M(\gamma_{H}\gamma_{L})-m(\pi^{0})|>0.015$~$\gevcc$ in the $\jpsi\to\uu$ mode. MC simulations show that the background from $\ee\to\gamma_{\rm ISR}\psi(3686)$ with $ \psi(3686)\to\gamma\chi_{c1,c2}$ can be ignored for most of the energies expect for data at $\sqrt{s}=3.773$~GeV. These background events are simulated and subtracted from the signal yield at $\sqrt{s}=3.773$~GeV. A fit to the lepton pair invariant mass gives a resolution of 10.8~$\mevcc$ and 10.5~$\mevcc$ for $\jpsi\to\ee$ and $\jpsi\to\mu^+\mu^-$ events, respectively. The $\jpsi$ mass window is defined as 3.08~$<M(\ell^+\ell^-)<3.12$~$\gevcc$. While the sidebands of the $\jpsi$ are defined by 3.00~$<M(\ell^+\ell^-)<3.06$~$\gevcc$ and 3.14~$<M(\ell^+\ell^-)<3.20$~$\gevcc$, which is three times as wide as the $\jpsi$ signal region.

\subsection{Cross section}
According to kinematic, we find that the photon from $\ee\to\gamma\chi_{c1,c2}$ has lower energy than the one from $\chi_{c1,c2}\to\gamma\jpsi$ for data with $\sqrt{s}<4.009$~GeV. On the contrary, the former has higher energy than the latter for data with $\sqrt{s}>4.009$~GeV. To obtain the number of signal events, we make use of both fitting and counting methods. For each data sample with $\cal{L}_{\rm int}>$ 400~pb$^{-1}$($\cal{L}_{\rm int}$ is integrated luminosity), a fit with $\jpsi\to\ee$ or $\uu$ events is performed to the invariant mass distribution of $\gamma_{H}\jpsi$ ($\sqrt{s}<4.009$~GeV) or $\gamma_{L}\jpsi$ ($\sqrt{s}>4.009$~GeV).  For data at $\sqrt{s}=4.009$~GeV, these two photons cannot be distinguished by energy. A 2-dimensional fit to the distribution of $M(\gamma_{H}\jpsi)$ versus $M(\gamma_{L}\jpsi)$ is used to extract the number of signal events. In these fits, the signal probability density functions (PDFs) are described with MC-simulated shapes, and the background PDFs are constrained to $\jpsi$ sideband events. For the low-statistics data samples with $\cal{L}_{\rm int}<$ 200~pb$^{-1}$ and $\sqrt{s}>4.009$~GeV, we obtain the signal yield by counting the number of events in the $\chi_{c1,c2}$ signal region and by subtracting the number of normalized background events in the $\chi_{c1,c2}$ mass sideband region. The $\chi_{c1}$ and $\chi_{c2}$ signal regions are defined as 3.49~$<M(\gamma_{L}\jpsi)<3.53$~$\gevcc$ and 3.54~$<M(\gamma_{L}\jpsi)<3.58$~$\gevcc$, which include more than 94$\%$ of the signal events. The sidebands of $\chi_{c1,c2}$ are defined as 3.42~$<M(\gamma_{L}\jpsi)<3.46$~$\gevcc$ and 3.6~$<M(\gamma_{L}\jpsi)<3.64$~$\gevcc$.

\begin{figure}[htp]
	\centering
	\subfigure{
	\includegraphics[width=3.2in,height=1.6in]{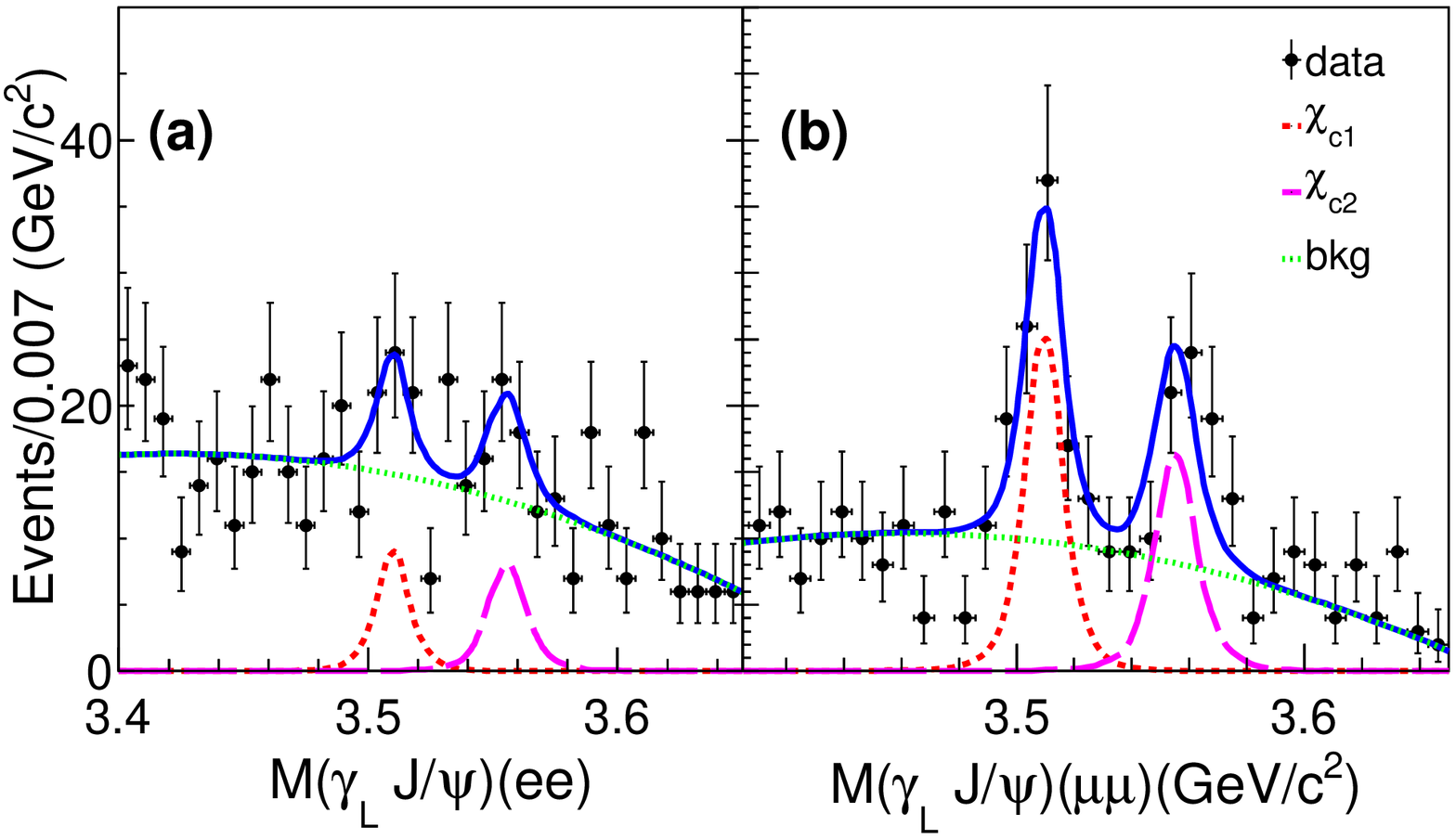}
	}

	\subfigure{
	\includegraphics[width=3.2in,height=1.6in]{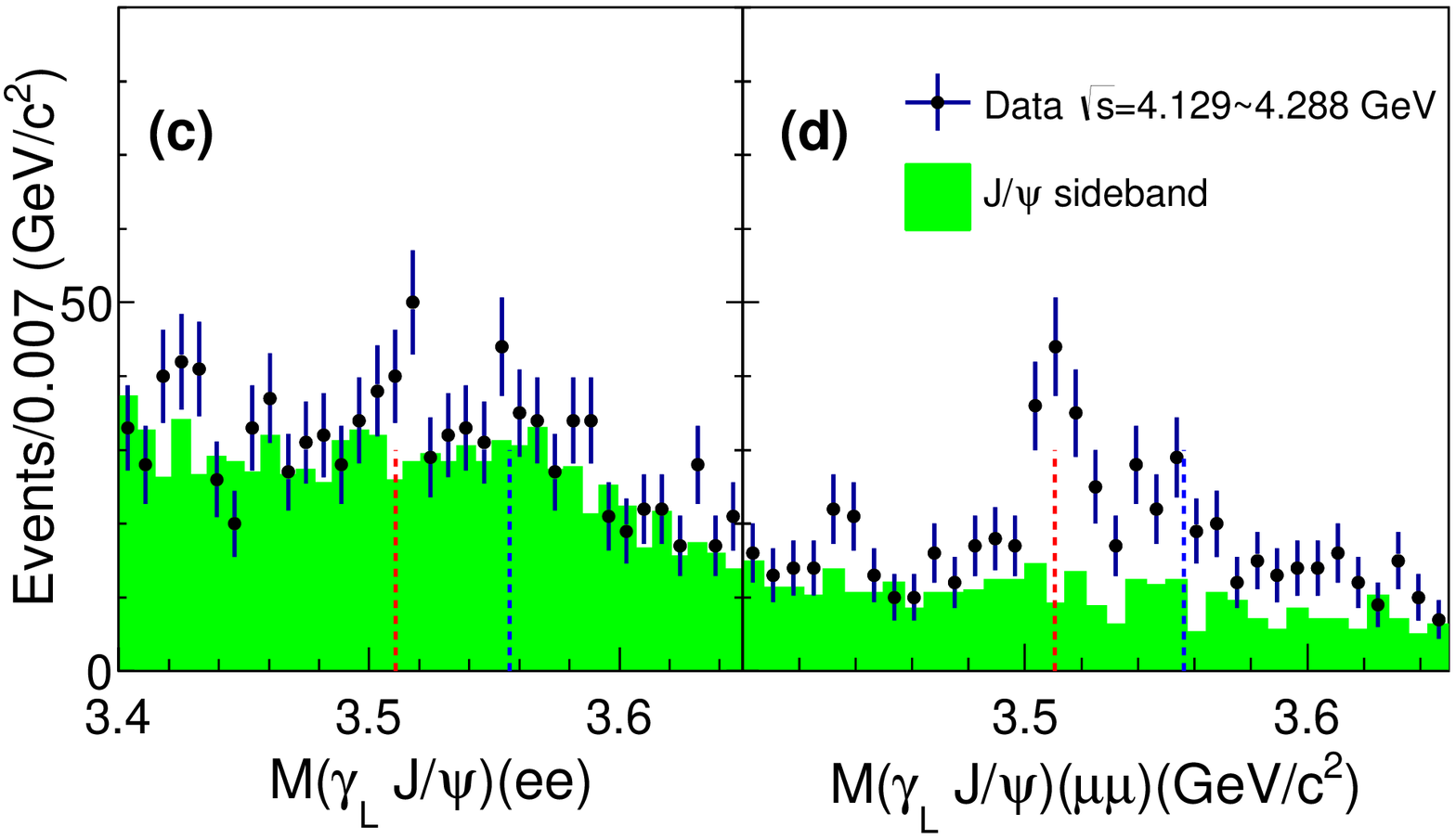}	
    }

\subfigure{                                 
	\includegraphics[width=3.2in,height=1.6in]{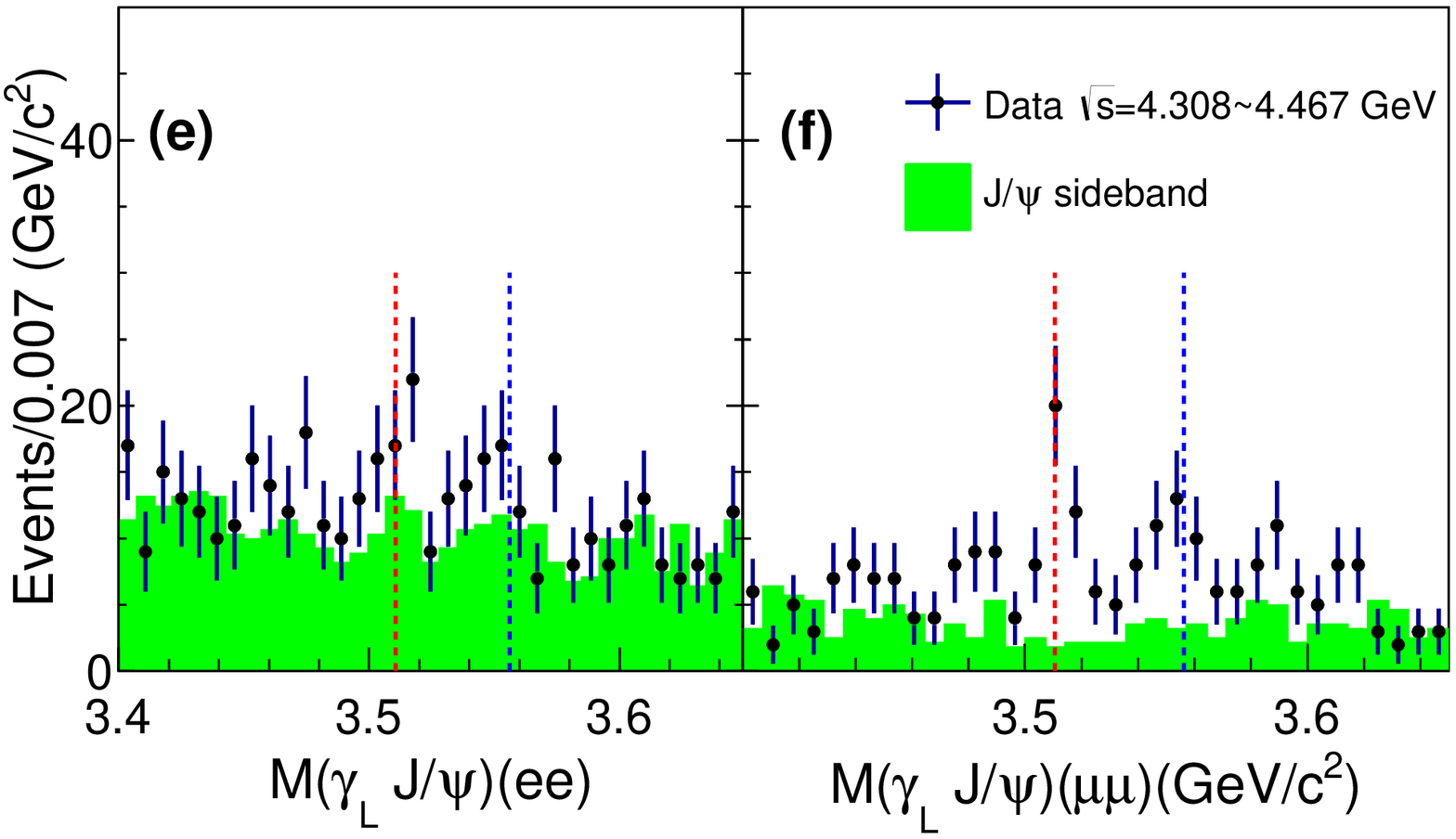}	
    }
	
	\caption{Fit to the $M(\gamma_{L}\jpsi)$ distributions for (a) $\jpsi\to\ee$ and (b) $\jpsi\to\uu$ data at $\sqrt{s}=4.178$~GeV. The $M(\gamma_{L}\jpsi)$ distributions for $\jpsi\to\ee$ (c,e) and $\jpsi\to\uu$ (d,f) data at $\sqrt{s}=4.129$ - $4.288$~GeV (exclude 4.178~GeV) (c,d) and  $\sqrt{s}=4.308$ - $4.467$~GeV (e,f). In a, b: Dots with error bars are data, the blue solid curves are the total fit results, the red dotted (pink dashed) curves are $\chi_{c1}$ ($\chi_{c2}$) signals, and the green dotted-dashed curves are backgrounds. In c, d, e, f: the shaded histograms are from normalized $\jpsi$ mass sideband events, the red and blue vertical dashed lines represent the world average mass values of $\chi_{c1}$ and $\chi_{c2}$ resonances, respectively.}
	\label{fig:chic12_4180}	
\end{figure}

Taking $\sqrt{s}=4.178$~GeV as an example, the invariant mass distribution of $M(\gamma_{L}\jpsi)$ as well as the fit results for the surviving events are shown in Fig.~\ref{fig:chic12_4180}. Clear $\chi_{c1,c2}$ signals are observed. The statistical significance of $\chi_{c1,c2}$ signals are calculated by comparing the log-likelihoods with and without the signal components in the fit, and taking the change of number of degrees of freedom into account. The statistical significances are estimated to be 7.6$\sigma$ for the $\chi_{c1}$ signal and 6.0$\sigma$ for the $\chi_{c2}$ signal. This is the first observation of the $\ee\to\gamma\chi_{c1,c2}$ processes between 4 - 5~GeV. The invariant mass distributions of $M(\gamma_{L}\jpsi)$ for both $\jpsi\to\ee$ and $\jpsi\to\uu$ at $\sqrt{s}=4.13\sim 4.3$~GeV  (exclude 4.178~GeV) and $\sqrt{s}=4.3\sim 4.5$~GeV are also shown in Fig.~\ref{fig:chic12_4180}.

\par
The production cross section of $\ee\to\gamma\chi_{c1,c2}$ at each $\ee$ c.m. energy is calculated as
\begin{linenomath}
\begin{equation}  
\sigma(\sqrt{s})=\frac{N^{\rm signal}}{\mathcal{L}_{\rm int}(1+\delta)\epsilon\mathcal{B}},
\end{equation}
\end{linenomath}
where $N^{\rm signal}$ is the number of signal events, $\mathcal{L}_{\rm int}$ is the integrated luminosity, $\epsilon$ is the selection efficiency, and $\mathcal{B}=\mathcal{B}(\chi_{c1,c2}\to\gamma\jpsi)\times\mathcal{B}(\jpsi\to\ell^{+}\ell^{-})$ is the branching fraction of intermediate states in the sequential decay, $(1+\delta)$ is the ISR correction factor~\cite{isr-cor}. 
The ISR correction factor is calculated with the {\sc kkmc} program, with the measured $\sqrt{s}$-dependent cross section of the reactions $\ee\to\gamma\chi_{c1,c2}$ as input.
This procedure is iterated several times until $(1+\delta)\epsilon$ converges, i.e. the relative difference between the last two iterations is less than 1\%.

The final measured cross sections $\sigma(\sqrt{s})$ for $\ee\to\gamma\chi_{c1}$ are shown in Fig.~\ref{fig:cross_c1} and are summarized 
in Table~\ref{tab:cross-db_c1} in the Appendix. Note that some of the cross sections are negative, thereby seem unphysical. This is caused by the fact that the number of events in the signal region is less than the estimated number of background events from sideband regions, and it can be explained by statistical fluctuations. To study the possible resonances in the $\ee\to\gamma\chi_{c1}$ process, a maximum likelihood fit is performed to the $\sqrt{s}$-dependent cross sections.
To describe the data, we use two coherent Breit-Wigner (BW) resonances, i.e. the $\psi(4040)$ and $\psi(4160)$, together with a continuum term and the incoherent $\psi(3686)$, $\psi(3770)$ tail
contributions. Since the contribution from $\psi(3686)$ and $\psi(3770)$ is small at $\sqrt{s}>4$~GeV and also lack of data between 3.77 and 4.0~GeV, we do not consider the interference effect from $\psi(3686)$ and $\psi(3770)$. The possible interference effect between continuum and other components is also investigated and we find its contribution is small (and taken as systematic effects).
The fit function is thus written as
\begin{linenomath}
\begin{equation}
\begin{split}
\sigma_{\ee\to\gamma\chi_{c1}}(\sqrt{s})=&|A_{cont}|^2+|BW_{\psi(3686)}(\sqrt{s})|^{2}\\
     &+|BW_{\psi(3770)}(\sqrt{s})|^2+    |BW_{\psi(4040)}(\sqrt{s})\\
     &+BW_{\psi(4160)}(\sqrt{s})e^{i\phi_1}|^{2},
\end{split} 
\end{equation}
\end{linenomath}
where $\phi_{i}$ is the relative phase of the amplitude, and $A_{cont}$ is the continuum amplitude
which is parametrized as
\begin{linenomath}
\begin{equation}
A_{cont}=\sqrt{\frac{f_{cont}}{\sqrt{s}^n}\Phi(\sqrt{s})},
\end{equation}
\end{linenomath}
where $f_{cont}$ and $n$ are the free parameters. $BW$ function is described as
\begin{linenomath}
\begin{equation}
BW_R(\sqrt{s})=\frac{M_R}{\sqrt{s}} \frac{\sqrt{12\pi\Gamma_R^{ee}\Gamma_R^{tot}{B_R}}}{s-M_R^{2}+iM_R\Gamma_R^{tot}}\sqrt{\frac{\Phi(\sqrt{s})}{\Phi(M_R)}},
\end{equation}
\end{linenomath}
where M, $\Gamma^{tot}_{R}$ and $\Gamma^{ee}_{R}$ are the mass, full width and electric width of the resonance $R$, respectively. ${B_{R}}$ is the branching fraction of $R\to\gamma\chi_{c1}$, and $\Phi(\sqrt{s})$ is the phase space factor. For the E1 transition of the process $\psi(3686)\to\gamma\chi_{c1}$, we consider an additional factor the $E_\gamma^3$~\cite{Brambilla:2010cs} and a damping factor~\cite{Anashin:2010dh}, according to 
\begin{linenomath}
\begin{equation}
BW_{\psi(3686)}(\sqrt{s})=\frac{M}{\sqrt{s}}\frac{\sqrt{12\pi\Gamma^{ee}\Gamma^{tot}B_{i}}}{s-M^{2}+iM\Gamma^{tot}}{\Phi(\sqrt{s}){D(\sqrt{s})}},
\end{equation}
\end{linenomath}
\\
where the phase space factor is given by ${\Phi(\sqrt{s})}=\left(\frac{E_\gamma}{E_\gamma^{0}}\right)^{3/2}$ ~\cite{Brambilla:2010cs}, and the damping factor as  $D(\sqrt{s})=\left(\frac{(E_\gamma^{0})^2}{E_\gamma^{0}E_{\gamma}+(E_\gamma^{0}-E_{\gamma})^2}\right)^{1/2} $ ~\cite{Anashin:2010dh}. The parameters $E_\gamma$ and $E_\gamma^{0}$ are the energy of the E1 photon for the $\psi(3686)\to\gamma{\chi_{c1}}$ decay at c.m. energy $\sqrt{s}$ and at the $\psi(3686)$ mass, respectively.

 In the cross section fit, the likelihood function is defined as 

\begin{linenomath}
	\begin{equation}
	\mathcal{L}=\prod {G}(N^{\rm sig}_{i}|s_i)\prod{P}(N^{\rm obs}_{j}|(s+b)_j),
	\end{equation}
\end{linenomath}

where $N^{\rm sig}$ is the number of signal events measured from dataset $i$, 
$s_i$ is the expected number of signal events for the corresponding dataset,
and $G$ represents a Gaussian distribution which describes datasets with high statistics
at $\sqrt{s}=3.773$ and $ 4.178$~GeV.
$N_j^{\rm obs}$ is the number of events observed in the $\chi_{c1}$ mass interval
from dataset $j$, $(s+b)_j$ is the expected sum of signal and background events
in the same interval, and $P$ represents a Poisson distribution which describes
low statistics datasets at other c.m. energies.
In the fit PDF, the masses and widths of $\psi(3686)$, $\psi(3773)$, $\psi(4040)$, $\psi(4160)$, and $\Gamma_{ee}\cdot\mathcal{B}_{i} [\psi(3686)\to\gamma\chi_{c1}]$ are fixed to PDG values~\cite{pdg}. The fit result is shown in Fig.~\ref{fig:cross_c1}, and also summarized in Table~\ref{tab:BW-fit_chic1}. The significance of $\psi(4040)$, $\psi(4160)$ and the continuum term are estimated to be 3.7$\sigma$, 3.3$\sigma$ and 6.7$\sigma$, respectively.  Considering the constructive and destructive interferences between $\psi(4040)$ and $\psi(4160)$,  there are two solutions with equal good quality from the fit, which has been proved mathematically~\cite{Bai:2019jrb}. A $\chi^{2}$-test is used to estimate the fit quality. Due to the low statistics of data at some c.m. energies, we merge the datasets into 17 groups, and the $\chi^{2}$-test gives $\chi^{2}/ndf = 10.6/11=0.96$, where $ndf$ is the number of degree of freedom.

\begin{figure}[htp]
	\centering
	\subfigure{
		\includegraphics[width=2.8in,height=2.3in]{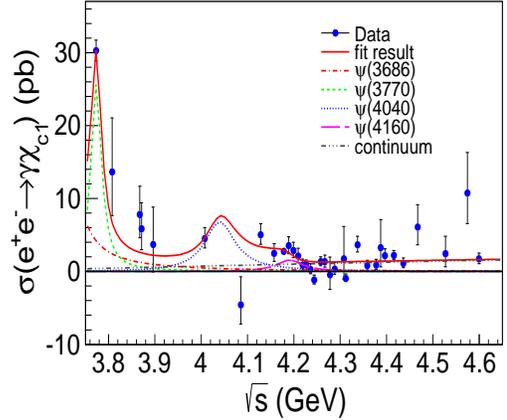}		
	}
	\caption{Cross section of $\ee\to\gamma\chi_{c1}$ process and a maximum likelihood fit to the line shape. Dots with error bars are data, the red curve shows
		the fit results, and the dashed curves show the contribution of each component.}
	\label{fig:cross_c1}	
\end{figure}

\begin{table}[htbp]
	\centering
	\caption{Results of the fit to the $\ee\to\gamma\chi_{c1}$ cross sections. The unit of $\ee$ partial width is~eV$/c^{2}$
		and the unit of $f_{cont}$ is ~eV$^n/$pb. 
		The errors are statistical only. }
	\label{tab:BW-fit_chic1}
	\begin{tabular}{c  r@{.}l  r@{.}l }
		\hline\hline
		\centering{Parameter} & \multicolumn{2}{c}{\centering{Solution I}}& \multicolumn{2}{c}{\centering{Solution II}} \\ 
		\hline
		$\Gamma^{ee}\mathcal{B}(\psi(3770)\to\gamma\chi_{c1})$ & \multicolumn{4}{c}{(6.8$\pm$0.4)$\times10^{-1}$}\\
		$\Gamma^{ee}\mathcal{B}(\psi(4040)\to\gamma\chi_{c1})$ & (6&0$\pm$2.1)$\times10^{-1}$&(6&1$\pm$2.1)$\times10^{-1}$ \\
		$\Gamma^{ee}\mathcal{B}(\psi(4160)\to\gamma\chi_{c1})$ & (1&3$\pm$0.8)$\times10^{-1}$&(1&4$\pm$0.9)$\times10^{-1}$\\		
		$\phi_1$ &192&$1^{\circ}\pm24.1^{\circ}$ &196&$0^{\circ}\pm24.6^{\circ}$  \\
		$f_{cont}$&\multicolumn{4}{c}{$4.1\pm0.6$}\\
		$n$ &\multicolumn{4}{c}{$0\pm1.3$}\\
		
		\hline\hline
	\end{tabular}
\end{table}

\begin{figure}[htp]
	\centering
	\subfigure{
		\includegraphics[width=2.8in,height=2.3in]{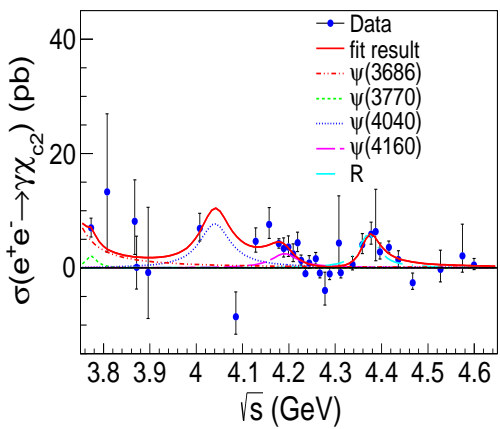}		
	}
	\caption{Cross section of $\ee\to\gamma\chi_{c2}$ process and a maximum likelihood fit to the line shape. Dots with error bars are data, the red curve shows
		the fit results, and the dashed curves show the contribution of each component.}
	\label{fig:cross_c2}	
\end{figure}

For the $\ee\to\gamma\chi_{c2}$ process, the measured cross sections are shown in the Fig.~\ref{fig:cross_c2} and summarized in Table~\ref{tab:cross-db_c2} in the Appendix. In the fit PDF, the resonance parameters of $\psi(3686)$, $\psi(3770)$, $\psi(4040)$,
and $\psi(4160)$ are also fixed to PDG values. To describe the $\sqrt{s}$-dependent cross section, one more resonance is added in the fit function to describe the 
structure around $\sqrt{s}=4.39$~GeV. According to the fit, the contribution from continuum is not significant in this process ($<1\sigma$). Thus, we construct the fit function as
\begin{linenomath}
	\begin{equation}
	\begin{split}
	\sigma_{\ee\to\gamma\chi_{c2}}(\sqrt{s})=&|BW_{\psi(3686)}(\sqrt{s})|^{2}+|BW_{\psi(3770)}(\sqrt{s})|^2+\\
	&|BW_{\psi(4040)}(\sqrt{s})+BW_{\psi(4160)}(\sqrt{s})e^{i\phi_1}\\
	&+BW_{\mathcal{R}}(\sqrt{s})e^{i\phi_2}|^{2}, 
	\end{split} 
	\end{equation} 
\end{linenomath}
The fit results are shown in Fig.~\ref{fig:cross_c2} and summarized in Table~\ref{tab:BW-fit_chic2}.
Same as before, there are four solutions with equal fit quality, due to the interferences between
$\psi(4040)$, $\psi(4160)$, and the new resonance~\cite{Bai:2019jrb}.  
The significance of $\psi(4040)$, $\psi(4160)$ and the resonance near 4.39~GeV 
are estimated to be 2.0$\sigma$, 4.6$\sigma$ and 5.8$\sigma$, respectively.
Similarly, we merge the data into 17 groups when performing a $\chi^{2}$-test. 
The $\chi^{2}$-test to the fit quality gives $\chi^{2}/ndf = 7.8/9=0.87$.

\begin{table*}
	\centering	
	\caption{Results of the fit to the $\ee\to\gamma\chi_{c2}$ cross sections. The unit of the $\ee$ partial width is~eV$/c^{2}$. The errors are statistical only.}	
\label{tab:BW-fit_chic2}	
	\begin{tabular}{c r@{.}l r@{.}l r@{.}l r@{.}l}	

		\hline\hline
		\centering{Parameter} & \multicolumn{2}{c}{\centering{Solution I}}& \multicolumn{2}{c}{\centering{Solution II}} & \multicolumn{2}{c}{\centering{Solution III}}& \multicolumn{2}{c}{\centering{Solution IV}}\\ 
		\hline
		$\Gamma^{ee}\mathcal{B}(\psi(3770)\to\gamma\chi_{c2})$ & \multicolumn{8}{c}{(0.6$\pm$0.4)$\times10^{-1}$}\\

		$\Gamma^{ee}\mathcal{B}(\psi(4040)\to\gamma\chi_{c2})$ & (13&4$\pm$4.7)$\times10^{-1}$&(6&9$\pm$3.5)$\times10^{-1}$ &(13&3$\pm$4.7)$\times10^{-1}$& (6&9$\pm$3.5)$\times10^{-1}$\\

		$\Gamma^{ee}\mathcal{B}(\psi(4160)\to\gamma\chi_{c2})$ & (6&8$\pm$1.9)$\times10^{-1}$&(2&1$\pm$0.9)$\times10^{-1}$ &(6&4$\pm$1.8)$\times10^{-1}$&(2&1$\pm$0.9)$\times10^{-1}$ \\
		$M(\mathcal{R})$ & 
		\multicolumn{8}{c}{4371.7$\pm7.5$}\\
		$\Gamma^{tot}(\mathcal{R})$ &
		\multicolumn{8}{c}{51.1$\pm17.6$} \\
		$\Gamma^{ee}\mathcal{B}(\mathcal{R}\to\gamma\chi_{c2})$ & 
		(4&7$\pm$1.6)$\times10^{-1}$&(3&9$\pm$1.3)$\times10^{-1}$ & (4&4$\pm$1.5)$\times10^{-1}$& (4&1$\pm$1.4)$\times10^{-1}$   \\
		$\phi_1$ &241&5$^{\circ}\pm15.0^{\circ}$&105&$6^{\circ}\pm33.7^{\circ}$ &  238&$9^{\circ}\pm14.8^{\circ}$ &107&$3^{\circ}\pm34.2^{\circ}$\\
		$\phi_2$ &248&$7^{\circ}\pm31.3^{\circ}$&24&$8^{\circ}\pm39.2^{\circ}$ & 252&$6^{\circ}\pm31.7^{\circ}$ &19&$5^{\circ}\pm30.8^{\circ}$\\
		\hline\hline		

	\end{tabular}

\end{table*}

\section{$\ee\to\gamma\chi_{c0}$}
\subsection{Event selection}
For the $\ee\to\gamma\chi_{c0}$ study, the $\chi_{c0}$ resonance is reconstructed with $2(\pi^{+}\pi^{-})$, $\pi^{+}\pi^{-}K^{+}K^{-}$, and $K^{+}K^{-}$ decay modes.
Considering the relatively small branching fractions from the $\chi_{c0}$ decay and also the 	high background levels, only the data samples with $L_{\rm int}>400$~pb$^{-1}$ at $\sqrt{s}>4.0$~GeV are used in this study. The selection criteria of charged tracks and photons are the same as for the $\ee\to\gamma\chi_{c1,c2}$ analysis. The particle identification (PID) of kaons and pions is based on the d$E/$d$x$ and TOF information, and the particle type with the highest probability is assigned to each track. For photons, the most energetic photon is regarded as the candidate for signal events. A 4C kinematic fit is performed to these three decay modes and $\chi_{4c}^{2}<25$ is required for both $\chi_{c0}\to2(\pi^+\pi^-)/ K^+K^-\pi^+\pi^-$ modes and $\chi_{4c}^{2}<30$ for the $\chi_{c0}\to K^+K^-$ mode. 

For the $\chi_{c0}\to K^+K^-\pi^+\pi^-$ decay mode, background events with a photon from resonances decay, such as $\omega\to\pi^{+}\pi^{-}\pi^{0}$, $\eta'\to\gamma\pi^{+}\pi^{-}$ and $\pi^{0}\to\gamma\gamma$ are vetoed. For $\omega\to\pi^{+}\pi^{-}\pi^{0}$ events with one of the photons from the $\pi^{0}$ decay undetected, we require $|M(\gamma\pi^+\pi^-)-756.9$~$\mevcc|>20$~$\mevcc$ to suppress them.  Here, $756.9$~$\mevcc$ is the position of the peak obtained by fitting the $M(\gamma\pi^+\pi^-)$ distribution
in the data, which has a $\sim$ 25~MeV mass shift from the $\omega$ world average mass~\cite{pdg}. 
The $\eta'\to\gamma\pi^{+}\pi^{-}$ background events are vetoed by requiring $|M(\gamma\pi^+\pi^-)-m(\eta')|>10$~$\mevcc$ (hereafter, $m$(particle) denotes the world average mass of a particle listed in the PDG~\cite{pdg}). To further suppress backgrounds from $\pi^{0}\to\gamma\gamma$ decay, the combination of the radiative photon with an extra reconstructed photon should not come from a $\pi^{0}$ candidate. We require $|M(\gamma\gamma_{extra})-m(\pi^{0})|>12$~$\mevcc$, where $M(\gamma\gamma_{extra})$ is the mass closest to $m(\pi^{0})$ from the radiative photon and an extra photon combination. Further background from $\phi\to K^{+}K^{-}$ process is also vetoed by requiring $M(K^{+}K^{-})>1.05$~$\gevcc$.

For the $\chi_{c0}\to 2(\pi^+\pi^-)$ decay mode, the background events with $\eta\to\gamma\pi^{+}\pi^{-}$, $\omega\to\pi^{0}\pi^{+}\pi^{-}$, $\eta'\to\gamma\pi^{+}\pi^{-}$ and $\pi^{0}\to\gamma\gamma$ are suppressed by requiring  $|M(\gamma\pi\pi)-m(\eta)|>6$~$\mevcc$, $|M(\gamma\pi\pi)-765.4$~$\mevcc|>22$~$\mevcc$, $|M(\gamma\pi\pi)-m(\eta')|>10$~$\mevcc$ and $|M(\gamma\gamma_{extra})-m(\pi^{0})|>6$~$\mevcc$, respectively. Similarly, the $765.4$~$\mevcc$ is the average value obtained by fitting the $M(\gamma\pi^+\pi^-)$ spectrum for $\omega$ background events. Here, $M(\gamma\pi\pi)$ keeps all combinations of pion pairs. There are backgrounds from radiative Bhabha and radiative dimuon events ($\ee\to\gamma\uu$), with one of the radiative photon converted to an $\ee$ pair ($\gamma$-conversion) and misidentified as pions. The opening angle of the $\pi^{+}\pi^{-}$ candidate is expected to be small ($\cos\theta\sim$1) for such kind of background events, and we require $\cos\theta_{\pi^{+}\pi^{-}}<0.98$ for all $\pi^+\pi^-$ candidate combinations to suppress them.

For the $\chi_{c0}\to K^+K^-$ decay mode, there are Bhabha background events. We require the deposited energy in the EMC over the momentum of a charged track $E_{EMC}/p<0.8$
to reject them. Same as before, $|M(\gamma\gamma_{extra})-m(\pi^{0})|>10$~$\mevcc$ is required to suppress background with $\pi^{0}\to\gamma\gamma$.

\subsection{Cross section}

\begin{figure*}[htp]
	\centering
	\subfigure{
		\includegraphics[width=2.0in,height=1.7in]{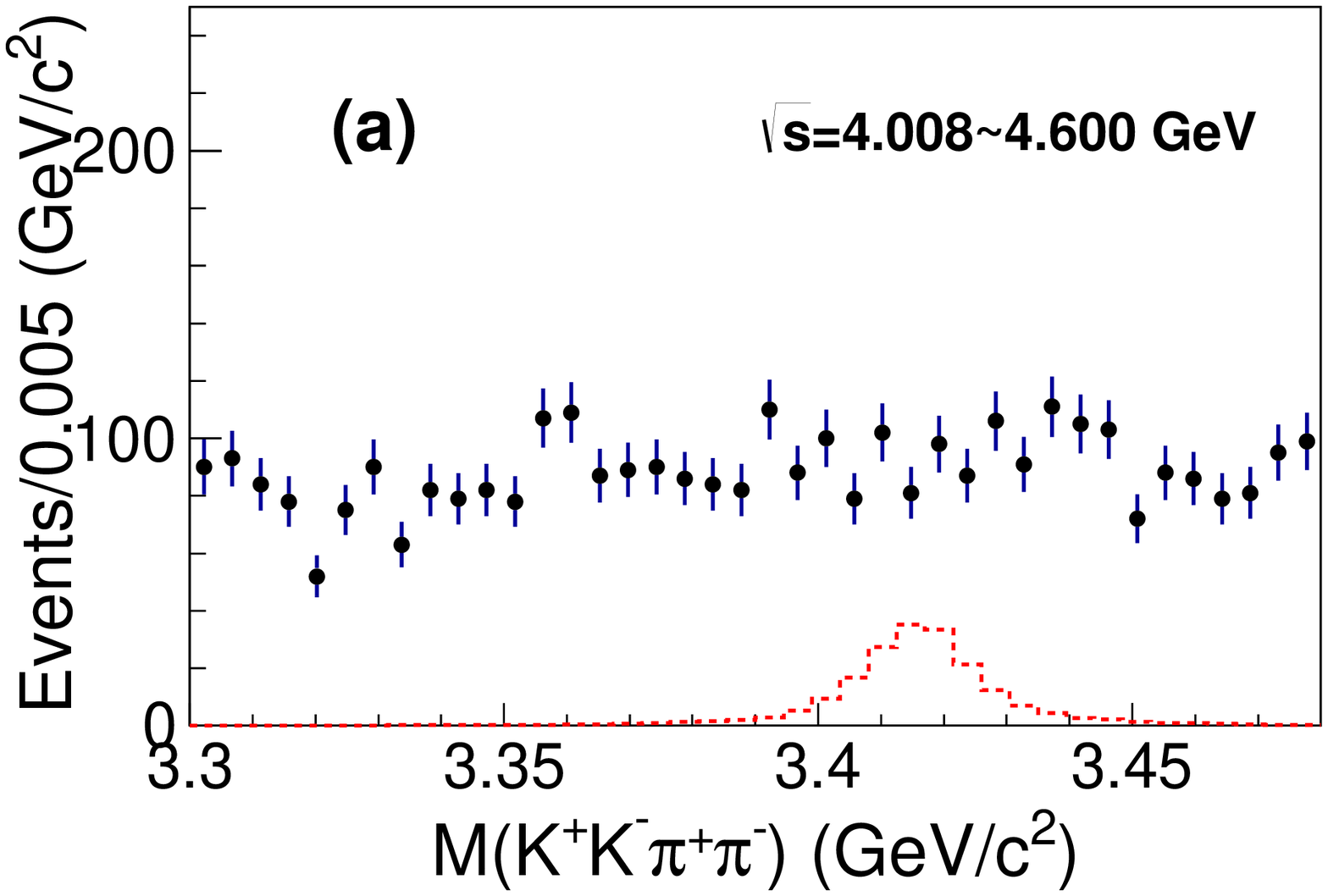}
		\includegraphics[width=2.0in,height=1.7in]{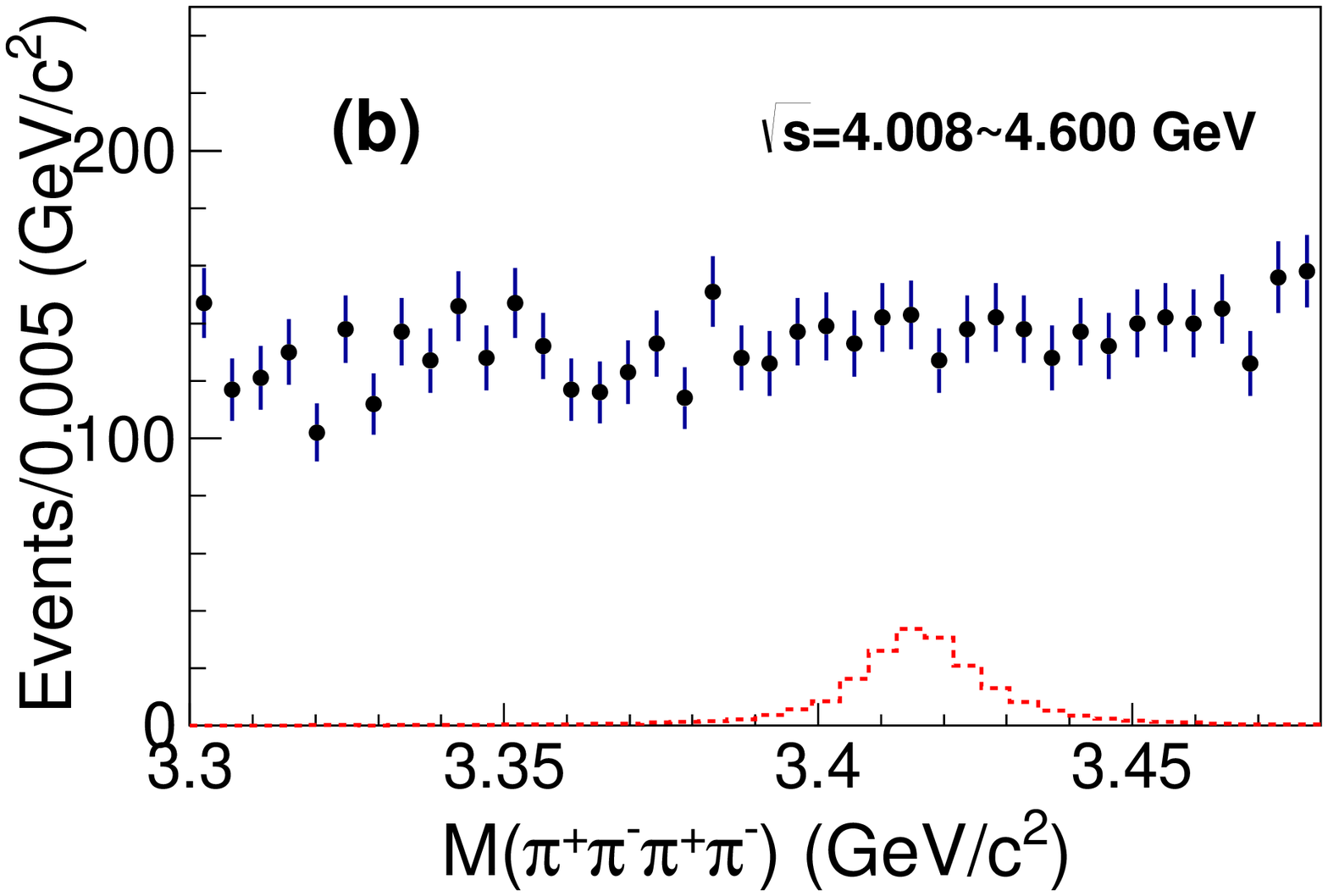}		
	}
	\subfigure{
		\includegraphics[width=2.0in,height=1.7in]{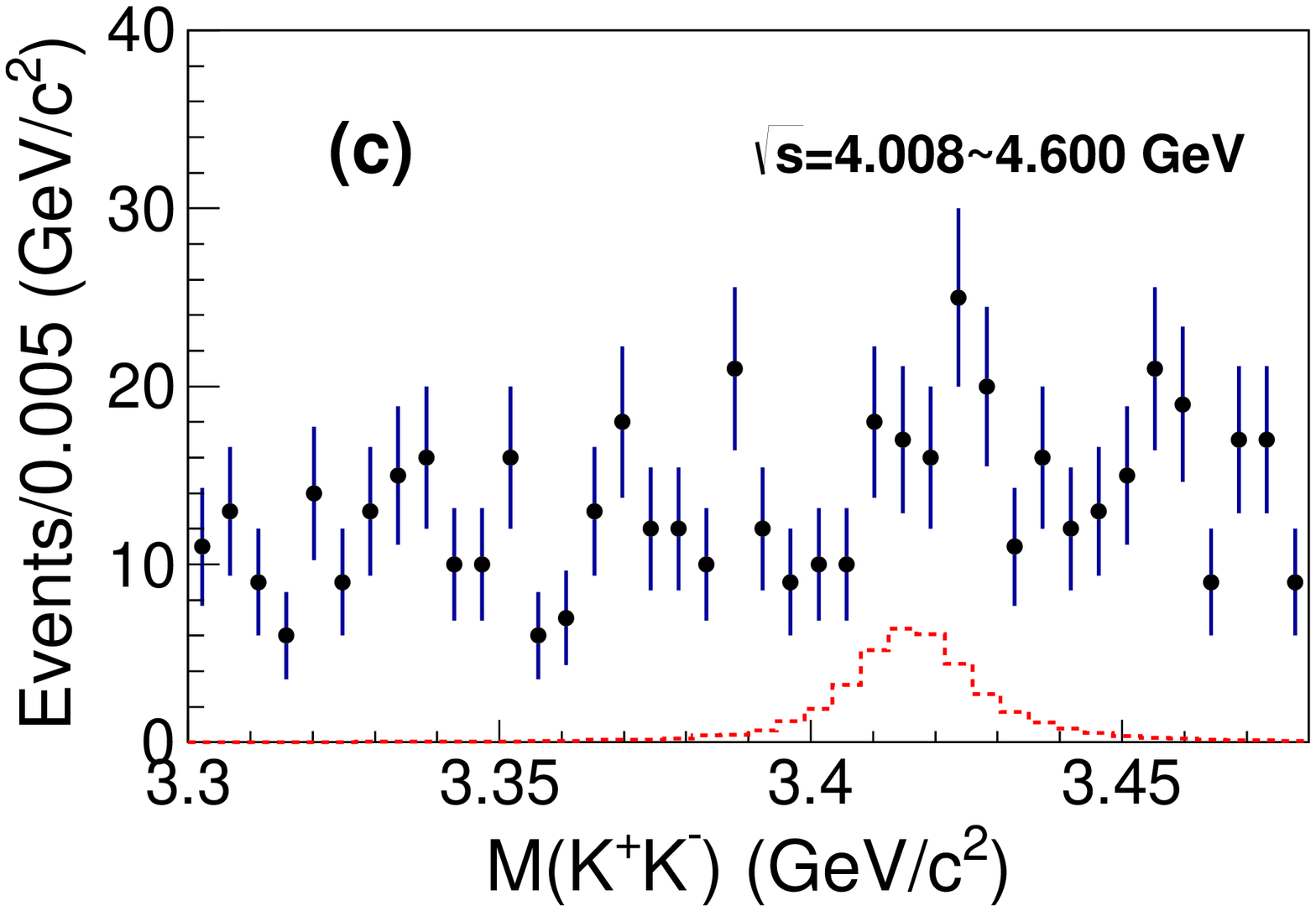}	
	}	
	\caption{Mass distributions of (a) $M(K^+K^-\pi^+\pi^-)$, (b) $M(\pi^+\pi^-\pi^+\pi^-)$ and (c) $M(K^+K^-)$ for combined data samples from $\sqrt{s}=4.008$ to 4.600~GeV. The red histograms represent the $\chi_{c0}$ MC shape with an arbitrary normalization.}
	\label{fig:data_c0}	
\end{figure*}

Figure~\ref{fig:data_c0} shows the $M(K^+K^-\pi^+\pi^-$), $M(\pi^+\pi^-\pi^+\pi^-$), and $M(K^+K^-)$ invariant mass distributions for the full datasets after imposing the 
above selection criteria. 
To obtain the number of $\chi_{c0}$ signal events, an unbinned maximum likelihood fit is performed to the $M(K^+K^-\pi^+\pi^-$), $M(\pi^+\pi^-\pi^+\pi^-$) and $M(K^+K^-)$ distributions simultaneously at each c.m. energy. The signal yields for three decay modes are constrained according to corresponding reconstruction efficiencies and branching fractions.
In the fit, the signal PDFs are described with the shapes from simulated signal MC events. The background shapes are described with two 2nd-order polynomial functions 
for the $\chi_{c0}\to K^+K^-\pi^+\pi^-$, $2(\pi^+\pi^-)$ decay modes, and a 1st-order polynomial function for the $K^+K^-$ mode. The significance of $\chi_{c0}$ signal is 
estimated to be less than 2$\sigma$ at each c.m. energy point. To estimate an upper limit (UL) of the production cross sections,
we scan the likelihood curve in the fit and set the 90\% C.L. The corresponding UL of the cross section is calculated as
\begin{linenomath}
	\begin{equation}  
	\sigma_{\ee\to\gamma\chi_{c0}}^{\rm up}(\sqrt{s})=\frac{N^{\rm up}}{\mathcal{L}_{\rm int}(1+\delta)(1+\delta_{v})\sum_{i=0}^{3}\epsilon_i\cdot\mathcal{B}_i}, 
	\end{equation}
\end{linenomath}
where $N^{\rm up}$ is the UL of the number of signal events at 90\% C.L., which is obtained by integrating the likelihood curve of the fit (the systematic uncertainty is considered
by convolving the likelihood curve with a Gaussian and its standard deviation is set to the systematic uncertainty). $(1+\delta_{v})$ is the vacuum polarization factor taken from calculation~\cite{VP-cite}. 
The UL of the cross sections at all c.m. energies are shown in Fig.~\ref{fig:UL-c0} and summarized in Table~\ref{tab:chi0_chic0_upper} in the Appendix .

\begin{figure}[htp]
	\centering
	
	\subfigure{
		\includegraphics[width=2.8in,height=2.3in]{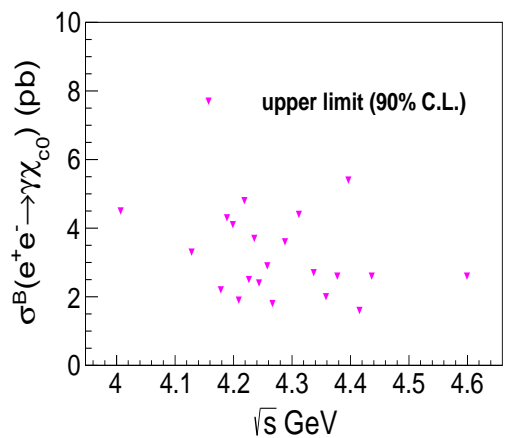}	
	}	
	\caption{The upper limits of Born cross section for $\ee\to\gamma\chi_{c0}$ process at $\sqrt{s}=4.008$ - $4.600$~GeV.}
	\label{fig:UL-c0}	
\end{figure}

\section{SYSTEMATIC UNCERTAINTY} 
The systematic uncertainty of the cross section measurements of $\ee\to\gamma\chi_{cJ}$ ($J=0,1,2$) mainly comes from the luminosity measurement, detection efficiency, decay branching fractions, signal extraction, and radiative correction. The luminosity is measured using Bhabha events and the uncertainty is estimated to be 1$\%$~\cite{Ablikim:2015nan}. For high momentum leptons, the uncertainty of the tracking efficiency is 1$\%$ per track~\cite{Ablikim:2014hwn}. The uncertainty in the photon reconstruction is 1$\%$ per photon, estimated by studying the $\jpsi\to\rho^{0}\pi^{0}$ decay~\cite{Ablikim:2010zn}. The PID efficiency uncertainty for each charged track is taken as 1$\%$~\cite{Ablikim:2017cbv}. For the systematic uncertainty from the kinematic fit, we correct the track helix parameters  in the MC simulation according to the method described in Ref.~\cite{Ablikim:2012pg}, and the efficiency difference before and after correction is considered as the systematic uncertainty. 
The uncertainties for the branching fractions of $\chi_{c1,c2}\to\gamma J/\psi$, and $\chi_{c0}\to\pi^+\pi^-\pi^+\pi^-,~K^+K^-\pi^+\pi^-,~K^+K^-$ from the PDG~\cite{pdg} are taken as systematic uncertainties for the cross section measurement. For the $\ee\to\gamma\chi_{c0}$ process, the systematic uncertainties for tracking, PID, photon detection and kinematic fit are the same, and the total systematic error is obtained by weighting each individual one according to the branching fractions and efficiencies of three $\chi_{c0}$ decay modes, by considering the possible correlations between them.

For the systematic uncertainty from the background veto requirements, we select the $\psi(3686)\to\gamma\chi_{c0,c1,c2}$ control samples, and the selection requirements are exactly the same as the requirements described above. We take the efficiency difference between MC simulation and corresponding control samples as the systematic uncertainties. For the systematic uncertainty from the $\jpsi$ mass window, an $\ee\to\eta\jpsi$ control sample is studied and we take the efficiency difference between MC simulation and control samples as the uncertainties.

\begin{table}[htbp]
	\centering
	\caption{Summary of systematic uncertainties sources on the cross section measurement in $\%$, the ``$-$" indicates that the uncertainty is not applicable.
	}
	\label{tab:sys_chic1c2}
	\begin{tabular}{c   |  c  |c |c  }
		\hline\hline
		\centering{Source} & {\centering{$\chi_{c0}$}} &  {\centering{$\chi_{c1}$}}& {\centering{$\chi_{c2}$}} \\ 
		\hline
		Luminosity  &  1.0 & 1.0&1.0\\
		Tracking & 3.7 & 2.0&2.0  \\
		Photon efficiency  & 1.0 & 2.0 &2.0 \\
		PID                & 3.7 & $-$ & $-$   \\ 
		Kinematic fit      & 2.4 & 0.3 &0.1\\		
		Branching fraction & 7.5 & 3.5 &3.6\\
		Signal extraction  & $-$ & 7.0 & 7.0\\
		Background veto    & 1.2 & 1.7 & 1.2 \\
		Decay model        & 1.3 & $-$ & $-$    \\
		Radiative correction & 1.2 & 4.5 & 3.9\\
		\hline
		Total & 9.8 & 9.7&9.4\\
		\hline\hline
	\end{tabular}
\end{table}

To obtain the systematic uncertainty from signal extraction, we refit the $M(\gamma_{H}\jpsi)$ by replacing the background shape with 2nd-order polynomial function, varying the fit range (+0.05~$\gevcc$), changing the signal shape from signal MC shape to signal MC shape convoluted with a float Gaussian function. 
The difference with nominal fit results is taken as the systematic uncertainty. For the systematic uncertainties from the ISR correction factor in $\ee\to\gamma\chi_{c1,c2}$, two sources are considered. First, the difference of $(1+\delta)*\epsilon$ between the last two iterations is taken as systematic uncertainty, which is 1\%. The other sources are the uncertainty of the fit parameters, fit components and damping factor. We sample the fit parameters with Gaussian functions (take the fit results as mean values and the errors as standard deviations) 200 times, then calculate the corresponding $(1+\delta)*\epsilon$ values. The standard deviation for $(1+\delta)*\epsilon$ of the 200 samplings is taken as a systematic error, which is 1.0\% and 1.8\% for $\ee\to\gamma\chi_{c1}$ and $\gamma\chi_{c2}$, respectively. Due to the low significance of $\psi(4040)$/$\psi(4160)$ in the $\ee\to\gamma\chi_{c1}$ process and $\psi(4040)$ in the $\ee\to\gamma\chi_{c2}$ process, the input cross section line shapes in MC generation are tested by excluding these charmonium states. The differences for $(1+\delta)*\epsilon$ are 3.5$\%$ and 3.0$\%$ for $\ee\to\gamma\chi_{c1}$ and
$\gamma\chi_{c2}$, respectively.  For the systematic uncertainty from the damping factor, we change the damping factor from $\left(\frac{(E_\gamma^{0})^2}{E_\gamma^{0}E_{\gamma}+(E_\gamma^{0}-E_{\gamma})^2}\right)^{1/2} $ ~\cite{Anashin:2010dh} to $e^{-\frac{E_\gamma^2}{8\beta^2}}$~\cite{Mitchell:2008aa}(the value of $\beta$ is also quoted from Ref.~\cite{Mitchell:2008aa}), the $(1+\delta)*\epsilon$ difference with two different damping factor are 0.9$\%$ and 1.3$\%$ for $\ee\to\gamma\chi_{c1}$ and $\gamma\chi_{c2}$, respectively. For the possible interference between continuum and other components in the $\ee\to\gamma\chi_{c1}$ process, the $(1+\delta)*\epsilon$ difference with or without considering  interference is taken as systematic error, which is 2.3$\%$. For the $\ee\to\gamma\chi_{c0}$, the difference between a flat line shape or a $\psi(3770)$ line shape is taken as uncertainty. For $\chi_{c0}\to\pi^+\pi^-\pi^+\pi^-, K^+K^-\pi^+\pi^-$ decay modes, the signal MC samples are generated by including all subprocesses. The difference with a pure phase space model is taken as the uncertainty due to the decay model. Table~\ref{tab:sys_chic1c2} summarizes all the systematic uncertainty sources and their contributions. The total systematic uncertainty is obtained by adding all sources in quadrature.

For the resonance parameters of the structure around 4.39~GeV, the uncertainty of c.m. energies ($\pm 0.8$~$\mev$) are common for all data samples, and this uncertainty will propagate directly to the mass measurement. For the uncertainty from the damping factor, we change the damping factor from  $\left(\frac{(E_\gamma^{0})^2}{E_\gamma^{0}E_{\gamma}+(E_\gamma^{0}-E_{\gamma})^2}\right)^{1/2} $ ~\cite{Anashin:2010dh} to $e^{-\frac{E_\gamma^2}{8\beta^2}}$~\cite{Mitchell:2008aa}, and the differences are 0.9~$\mev$ and 0.5~$\mev$ for the mass and width, respectively. To estimate the uncertainties from the parameters of $\psi(3686), \psi(3770), \psi(4160)$ and $\psi(4040)$, we randomly sample the four resonances parameters with Gaussian functions (PDG means and errors), and these values are used as input to refit the cross section. The standard deviations of these 1500 fit results are quoted as systematic errors, which are 1.0 $\mevcc$ and 1.6 $\mev$ for the mass and width, respectively. In the cross section fit, the $\psi(3770)$ contribution is added incoherently in the PDF. The possible systematic from the interference effect of $\psi(3770)$ is estimated by considering interference effect in the cross section fit. The differences are 0.8~$\mevcc$ and 0.8~$\mev$ for the mass and width. Assuming all the systematic errors are independent, the total systematic errors are 1.8~$\mevcc$ and 1.9~$\mev$ for mass and width, respectively, by adding all sources in quadrature.

\section{SUMMARY}    
In summary, using 19.3 fb$^{-1}$ data at c.m. energies between 3.773 and  4.600~GeV, we observe the $\ee\to\gamma\chi_{c1,c2}$ processes for the first time at $\sqrt{s}=4.178$~GeV. The statistical significances are 7.6$\sigma$ and 6.0$\sigma$ for $\gamma\chi_{c1,c2}$, respectively. For the $\ee\to\gamma\chi_{c1}$ process, the cross section line shape can be described with $\psi(3686)$, $\psi(3770)$, $\psi(4040)$ and $\psi(4160)$ resonances. For the $\ee\to\gamma{\chi_{c2}}$ process, one more resonance is added to describe the line shape of the cross section. The
significance of this resonance is estimated to be $5.8\sigma$, and its parameters are measured to be $M=4371.7\pm7.5\pm1.8$~$\mevcc$ and $\Gamma^{tot}=51.1\pm17.6\pm1.9$~$\mev$, which are consistent with the $Y(4360)$/$Y(4390)$ resonances~\cite{pdg} within errors. Our result supports the $Y(4360)/Y(4390)\to\gamma\chi_{c2}$ radiative transition.
In addition, the measured cross sections for $\ee\to\gamma\chi_{c1,c2}$ are consistent with the potential model predictions~\cite{Barnes:2005pb}, except for $\mathcal{B}[\psi(4160)\to\gamma\chi_{c2}]\sim10^{-7}$, which is significantly lower than our measurement $\mathcal{B}[\psi(4160)\to\gamma\chi_{c2}]$=(4.4 - 14.2$)\times 10^{-4}$.
For the $\ee\to\gamma\chi_{c0}$ process, no obvious signal is observed. The UL indicates the $\ee\to\gamma\chi_{c0}$ cross section is less than 8~pb between 4 and 4.6~GeV, and the UL is consistent with theoretical expectations.

 \section{ACKNOWLEDGEMENTS}
The BESIII collaboration thanks the staff of BEPCII and the IHEP computing center for their strong support. This work is supported in part by National Key R$\&$D Program of China under Contracts No. 2020YFA0406300, No. 2020YFA0406400; National Natural Science Foundation of China (NSFC) under Contracts No. 11625523, No. 11635010, No. 11735014, No. 11822506, No. 11835012, No. 11935015, No. 11935016, No. 11935018, No. 11961141012, No. 12022510, No. 12025502, No. 12035009, No. 12035013, No. 12061131003, No. 11975141; the Chinese Academy of Sciences (CAS) Large-Scale Scientific Facility Program; Joint Large-Scale Scientific Facility Funds of the NSFC and CAS under Contracts No. U1732263, No. U1832207; CAS Key Research Program of Frontier Sciences under Contract No. QYZDJ-SSW-SLH040; 100 Talents Program of CAS; INPAC and Shanghai Key Laboratory for Particle Physics and Cosmology; ERC under Contract No. 758462; European Union Horizon 2020 research and innovation programme under Contract No. Marie Sklodowska-Curie grant agreement No 894790; German Research Foundation DFG under Contracts Nos. 443159800, Collaborative Research Center CRC 1044, FOR 2359, FOR 2359, GRK 214; Istituto Nazionale di Fisica Nucleare, Italy; Ministry of Development of Turkey under Contract No. DPT2006K-120470; National Science and Technology fund; Olle Engkvist Foundation under Contract No. 200-0605; STFC (United Kingdom); The Knut and Alice Wallenberg Foundation (Sweden) under Contract No. 2016.0157; The Royal Society, UK under Contracts No. DH140054, No. DH160214; The Swedish Research Council; U. S. Department of Energy under Contracts No. DE-FG02-05ER41374, No. DE-SC-0012069


\onecolumngrid	
\begin{appendices}
\newpage	
\section{Appendix}

\setcounter{table}{0}  
\renewcommand\thetable{\Alph{table}}
\begin{table*}[htpb]
	\footnotesize
	\centering
	\caption{Summary of the c.m. energy, luminosities, the number of signals, detection efficiencies($\ee$ and $\uu$ mode), radiative correction factors, measured cross section($\sigma(\ee\to\gamma\chi_{c1})$). The first errors are statistical and the second systematic. The ``$-$" indicates that when we measure cross section, we merge $\jpsi\to\ee$ and $\jpsi\to\uu$ two modes, and only combined cross section is obtained.}
	\label{tab:cross-db_c1}
	\begin{tabular}{ c |  c|  r@{.}l|r@{.}l | c | c | c | r@{.}l|r@{.}l| r@{.}l }
		\hline\hline
		\centering{$\sqrt{s}$} & \centering{$\mathcal{L}_{\rm int}$(pb$^{-1}$)} &
		\multicolumn{2}{c|}{\centering{$N_{c1(\ee)}$}}&\multicolumn{2}{c|}{\centering{$N_{c1(\uu)}$}} &\centering{$\epsilon_{e^{+}e^{-}}$} & \centering{$\epsilon_{\mu^{+}\mu^{-}}$} & \centering{$(1+\delta)$} &\multicolumn{2}{c|}{$\sigma(e^{+}e^{-})$~(pb)}&    \multicolumn{2}{c|}{$\sigma(\mu^{+}\mu^{-})$~(pb)}& \multicolumn{2}{c}{$\sigma_{\rm com}$~(pb)}\\ 
		\hline
		3.7730 & 2932 & 281&$7^{+23.3}_{-23.1}$ & 450&$0^{+26.2}_{-25.4}$ & 0.210 & 0.369 & 0.689& 32&$4^{+2.7}_{-2.7}\pm3.2$&29&$5^{+1.7}_{-1.7}\pm2.8$ & 30&$3^{+1.4}_{-1.4}\pm2.9$  \\
		3.8077 & 50.5 & \multicolumn{4}{c|}{$7.0^{+3.8}_{-3.1}$} & \multicolumn{2}{c|}{0.252} & 0.983&\multicolumn{2}{c|}{-}&\multicolumn{2}{c|}{-}&13&$6^{+7.4}_{-6.0}\pm1.3$ \\
		3.8675 & 109 &\multicolumn{4}{c|}{ $8.0^{+4.0}_{-3.3}$} & \multicolumn{2}{c|}{0.207} & 1.114&\multicolumn{2}{c|}{-}&\multicolumn{2}{c|}{-}&7&$8^{+3.9}_{-3.2}\pm0.8$ \\
		3.8715 & 110 & \multicolumn{4}{c|}{$6.1^{+3.7}_{-3.0}$} & \multicolumn{2}{c|}{0.207} & 1.113&\multicolumn{2}{c|}{-}&\multicolumn{2}{c|}{-}&5&$8^{+3.6}_{-2.9}\pm0.6$ \\
		3.8962 & 52.6 & \multicolumn{4}{c|}{$1.8^{+2.5}_{-1.9}$ }& \multicolumn{2}{c|}{0.208} & 1.105&\multicolumn{2}{c|}{-}&\multicolumn{2}{c|}{-}&3&$7^{+5.1}_{-3.7}\pm0.4$ \\	
		4.0076 & 482 & 8&$6^{+5.1}_{-4.3}$ & 12&$7^{+5.1}_{-4.3}$ & 0.194 & 0.358 & 0.849& 5&$3^{+3.1}_{-2.6}\pm0.5$&4&$3^{+1.7}_{-1.5}\pm0.4$ & 4&$5^{+1.5}_{-1.3}\pm0.4$ \\
		4.0855 & 52.9 &\multicolumn{4}{c|}{ $-2.5^{+2.1}_{-1.4}$} & \multicolumn{2}{c|}{0.269} & 0.948&\multicolumn{2}{c|}{-}&\multicolumn{2}{c|}{-}&-4&$6^{+3.9}_{-2.6}\pm0.3$ \\		
		4.1285 & 394 & 5&$9^{+4.3}_{-3.5}$ & 14&$3^{+4.7}_{-4.0}$ & 0.175 & 0.321 & 1.027& 4&$1^{+2.9}_{-2.4}\pm0.4$&5&$4^{+1.8}_{-1.5}\pm0.5$ & 5&$0^{+1.5}_{-1.3}\pm0.5$  \\
		4.1574 & 407 & 4&$3^{+4.4}_{-3.6}$ & 6&$4^{+4.0}_{-3.3}$ & 0.171 & 0.318 & 1.021& 2&$9^{+3.0}_{-2.5}\pm0.3$&2&$4^{+1.5}_{-1.2}\pm0.2$ & 2&$5^{+1.3}_{-1.1}\pm0.2$\\
		4.1783 & 3189 & 23&$0^{+10.1}_{-9.4}$ & 65&$3^{+11.4}_{-10.8}$ & 0.173 & 0.322 & 1.031& 2&$0^{+0.9}_{-0.8}\pm0.2$&3&$0^{+0.5}_{-0.5}\pm0.3$ & 2&$7^{+0.5}_{-0.4}\pm0.3$  \\
		4.1888 & 566 & -0&$7^{+5.1}_{-4.4}$ & 18&$3^{+5.4}_{-4.8}$ & 0.164 & 0.309 & 1.075& -0&$3^{+2.5}_{-2.2}\pm0.1$&4&$8^{+1.4}_{-1.2}\pm0.5$ & 3&$5^{+1.2}_{-1.1}\pm0.3$ \\
		4.1989 & 526 & 2&$2^{+3.9}_{-3.2}$ & 13&$9^{+5.2}_{-4.5}$ & 0.159 & 0.297 & 1.146& 1&$1^{+2.0}_{-1.6}\pm0.1$&3&$8^{+1.4}_{-1.2}\pm0.4$ & 2&$9^{+1.2}_{-1.0}\pm0.3$ \\
		4.2092 & 517 & 3&$7^{+4.7}_{-4.0}$ & 7&$9^{+4.4}_{-3.6}$ & 0.147 & 0.282 & 1.220& 2&$0^{+2.5}_{-2.1}\pm0.2$&2&$2^{+1.2}_{-1.0}\pm0.2$ & 2&$1^{+1.1}_{-0.9}\pm0.2$ \\
		4.2187 & 515 & 1&$1^{+4.4}_{-3.7}$ & 3&$3^{+3.4}_{-2.7}$ & 0.146 & 0.276 & 1.269& 0&$6^{+2.3}_{-1.9}\pm0.1$&0&$9^{+0.9}_{-0.7}\pm0.1$ & 0&$9^{+0.9}_{-0.7}\pm0.1$  \\
		4.2263 & 1101 & 3&$9^{+5.8}_{-5.1}$ & 7&$6^{+5.0}_{-4.3}$ & 0.144 & 0.265 & 1.290& 0&$9^{+1.4}_{-1.2}\pm0.1$&1&$0^{+0.7}_{-0.6}\pm0.1$ & 1&$0^{+0.6}_{-0.5}\pm0.1$ \\
		4.2357 & 530 & 0&$5^{+3.8}_{-3.1}$ & 1&$1^{+3.1}_{-2.4}$ & 0.142 & 0.264 & 1.296& 0&$3^{+1.9}_{-1.5}\pm0.1$&0&$3^{+0.8}_{-0.7}\pm0.1$ & 0&$3^{+0.8}_{-0.6}\pm0.1$  \\
		4.2438 & 538 & 3&$6^{+4.7}_{-3.9}$ & -5&$4^{+2.6}_{-2.2}$ & 0.142 & 0.264 & 1.289& 1&$8^{+2.3}_{-2.0}\pm0.2$&-1&$4^{+0.7}_{-0.6}\pm0.1$ & -1&$2^{+0.7}_{-0.6}\pm0.1$  \\
		4.2580 & 828 & 13&$2^{+5.8}_{-5.0}$ & 5&$4^{+4.0}_{-3.3}$ & 0.143 & 0.263 & 1.265& 4&$3^{+1.9}_{-1.6}\pm0.4$&1&$0^{+0.7}_{-0.6}\pm0.1$ & 1&$4^{+0.7}_{-0.5}\pm0.1$\\
		4.2668 & 531 & 1&$6^{+3.4}_{-2.7}$ & 5&$6^{+3.9}_{-3.2}$ & 0.147 & 0.262 & 1.250& 0&$8^{+1.7}_{-1.3}\pm0.1$&1&$6^{+1.1}_{-0.9}\pm0.2$ & 1&$3^{+0.9}_{-0.7}\pm0.1$  \\
		4.2778 & 176 & \multicolumn{4}{c|}{$-0.8^{+4.4}_{-3.6}$} & \multicolumn{2}{c|}{0.203} & 1.230&\multicolumn{2}{c|}{-}&\multicolumn{2}{c|}{-}&-0&$5^{+2.4}_{-2.0}\pm0.1$ \\		
		4.2879 & 492 & -1&$4^{+3.0}_{-2.2}$ & 2&$8^{+3.6}_{-2.9}$ & 0.148 & 0.268 & 1.225& -0&$8^{+1.6}_{-1.2}\pm0.1$&0&$8^{+1.1}_{-0.9}\pm0.1$ & 0&$3^{+0.9}_{-0.7}\pm0.1$  \\
		4.3079 & 45.1 &\multicolumn{4}{c|}{ $0.8^{+2.1}_{-1.3}$} & \multicolumn{2}{c|}{0.212} & 1.189&\multicolumn{2}{c|}{-}&\multicolumn{2}{c|}{-}& 1&$7^{+4.4}_{-2.8}\pm0.1$\\
		4.3121 & 492 & 0&$3^{+3.3}_{-2.5}$ & -3&$7^{+2.2}_{-1.8}$ & 0.152 & 0.276 & 1.189& 0&$2^{+1.8}_{-1.4}\pm0.1$&-1&$1^{+0.7}_{-0.5}\pm0.1$ & -1&$0^{+0.6}_{-0.5}\pm0.1$  \\
		4.3374 & 501 & 7&$1^{+4.6}_{-3.9}$ & 12&$2^{+4.4}_{-3.8}$ & 0.156 & 0.286 & 1.162& 3&$8^{+2.5}_{-2.1}\pm0.4$&3&$6^{+1.3}_{-1.1}\pm0.3$ & 3&$7^{+1.2}_{-1.0}\pm0.4$  \\
		4.3583 & 544 & -1&$3^{+3.5}_{-2.7}$ & 4&$3^{+3.2}_{-2.5}$ & 0.163 & 0.293 & 1.142& -0&$6^{+1.7}_{-1.3}\pm0.1$&1&$2^{+0.9}_{-0.7}\pm0.1$ & 0&$8^{+0.8}_{-0.6}\pm0.1$  \\
		4.3774 & 523 &9&$4^{+4.8}_{-4.1}$ & 1&$8^{+2.9}_{-2.2}$ & 0.160 & 0.297 & 1.129& 4&$9^{+2.5}_{-2.1}\pm0.5$&0&$5^{+0.8}_{-0.6}\pm0.1$ & 0&$9^{+0.8}_{-0.6}\pm0.1$  \\
		4.3874 & 55.6 &\multicolumn{4}{c|}{ $1.9^{+2.2}_{-1.5}$} & \multicolumn{2}{c|}{0.225} & 1.124&\multicolumn{2}{c|}{-}&\multicolumn{2}{c|}{-}&3&$3^{+3.9}_{-2.6}\pm0.2$ \\		
		4.3965 & 505 & 3&$0^{+3.6}_{-2.8}$ & 8&$0^{+3.7}_{-3.0}$ & 0.159 & 0.299 & 1.117& 1&$7^{+1.9}_{-1.5}\pm0.2$&2&$3^{+1.1}_{-0.9}\pm0.2$ & 2&$2^{+1.0}_{-0.8}\pm0.2$  \\
		4.4156 & 1091 & 9&$7^{+5.3}_{-4.6}$ & 16&$2^{+5.4}_{-4.7}$ & 0.165 & 0.304 & 1.107& 2&$4^{+1.3}_{-1.1}\pm0.2$&2&$2^{+0.7}_{-0.6}\pm0.2$ & 2&$2^{+0.6}_{-0.6}\pm0.2$ \\
		4.4362 & 568 & 3&$3^{+3.9}_{-3.2}$ & 3&$8^{+3.2}_{-2.4}$ & 0.168 & 0.304 & 1.092& 1&$5^{+1.8}_{-1.5}\pm0.1$&1&$0^{+0.8}_{-0.6}\pm0.1$ & 1&$1^{+0.8}_{-0.6}\pm0.1$  \\
		4.4671 & 111 & \multicolumn{4}{c|}{$7.0^{+3.5}_{-2.8}$} & \multicolumn{2}{c|}{0.235} & 1.084&\multicolumn{2}{c|}{-}&\multicolumn{2}{c|}{-}&6&$1^{+3.0}_{-2.4}\pm0.4$ \\		
		4.5271 & 112 & \multicolumn{4}{c|}{$2.9^{+2.8}_{-2.1}$} &\multicolumn{2}{c|}{ 0.239} & 1.066&\multicolumn{2}{c|}{-}&\multicolumn{2}{c|}{-}&2&$4^{+2.4}_{-1.8}\pm0.2$ \\  
		4.5745 & 48.9 & \multicolumn{4}{c|}{$5.5^{+2.9}_{-2.2}$} & \multicolumn{2}{c|}{0.244} & 1.053&\multicolumn{2}{c|}{-}&\multicolumn{2}{c|}{-}&10&$7^{+5.6}_{-4.2}\pm0.7$ \\
		4.5995 & 587 & 3&$0^{+3.8}_{-3.0}$ & 7&$4^{+3.6}_{-3.0}$ & 0.172 & 0.324 & 1.047& 1&$4^{+1.8}_{-1.4}\pm0.1$&1&$8^{+0.9}_{-0.7}\pm0.2$ & 1&$7^{+0.8}_{-0.6}\pm0.2$ \\
		\hline\hline
	\end{tabular}
\end{table*}

\newpage
\begin{table*}[htp]
	\footnotesize
	\centering
	\caption{Summary of the c.m. energy, luminosities, detection efficiencies($\ee$ and $\uu$ mode), radiative correction factors, measured cross section($\sigma(\ee\to\gamma\chi_{c2})$). The first errors are statistical and the second systematic.}
	\label{tab:cross-db_c2}
	\begin{tabular}{ c |  c|  r@{.}l|r@{.}l | c | c | c | r@{.}l|r@{.}l| r@{.}l}
		\hline\hline
		\centering{$\sqrt{s}$} & \centering{$\mathcal{L}_{\rm int}$(pb$^{-1}$)} &
		\multicolumn{2}{c|}{\centering{$N_{c2(\ee)}$}}&\multicolumn{2}{c|}{\centering{$N_{c2(\uu)}$}} &\centering{$\epsilon_{e^{+}e^{-}}$} & \centering{$\epsilon_{\mu^{+}\mu^{-}}$} & \centering{$(1+\delta)$} &\multicolumn{2}{c|}{$\sigma(e^{+}e^{-})$~(pb)}&    \multicolumn{2}{c|}{$\sigma(\mu^{+}\mu^{-})$~(pb)}& \multicolumn{2}{c}{$\sigma_{\rm com}$~(pb)}\\ 
		\hline
		3.7730 & 2932 & 26&$5^{+17.8}_{-17.1}$ & 60&$9^{+15.6}_{-15.0}$ & 0.191 & 0.334 & 0.746& 5&$6^{+3.8}_{-3.6}\pm0.5$&7&$4^{+1.9}_{-1.8}\pm0.7$ & 7&$0^{+1.7}_{-1.6}\pm0.7$ \\
		
		3.8077 & 50.5 & \multicolumn{4}{c|}{$3.1^{+3.2}_{-2.5}$} & \multicolumn{2}{c|}{0.233} & 0.877&\multicolumn{2}{c|}{-}&\multicolumn{2}{c|}{-}& 13&$3^{+13.6}_{-10.6}\pm1.3$ \\
		3.8675 & 109 & \multicolumn{4}{c|}{$4.2^{+3.7}_{-3.0}$} & \multicolumn{2}{c|}{0.220} & 0.946&\multicolumn{2}{c|}{-}&\multicolumn{2}{c|}{-}&8&$1^{+7.3}_{-5.8}\pm0.8$ \\
		3.8715 & 110 &\multicolumn{4}{c|}{$0.0^{+2.8}_{-2.0}$} & \multicolumn{2}{c|}{0.222} & 0.945&\multicolumn{2}{c|}{-}&\multicolumn{2}{c|}{-}&0&$1^{+5.4}_{-3.8}\pm0.1$ \\
		3.8962 & 52.6 &\multicolumn{4}{c|}{ $-0.2^{+2.9}_{-2.1}$} & \multicolumn{2}{c|}{0.228} & 0.942&\multicolumn{2}{c|}{-}&\multicolumn{2}{c|}{-}&-0&$8^{+11.4}_{-8.0}\pm0.1$ \\

		4.0076 & 482 & 12&$1^{+5.4}_{-4.7}$ & 8&$4^{+4.5}_{-3.8}$ & 0.201 & 0.364 & 0.767& 14&$1^{+6.4}_{-5.5}\pm1.3$&5&$4^{+2.9}_{-2.5}\pm0.5$ & 6&$9^{+2.7}_{-2.2}\pm0.6$  \\
		4.0855 & 52.9 & \multicolumn{4}{c|}{$-2.4^{+1.2}_{-0.9}$} & \multicolumn{2}{c|}{0.253} & 0.936&\multicolumn{2}{c|}{-}&\multicolumn{2}{c|}{-}& -8&$6^{+4.4}_{-3.0}\pm0.6$  \\		
		4.1285 & 394 & 2&$8^{+4.0}_{-3.2}$ & 6&$9^{+3.7}_{-3.0}$ & 0.166 & 0.313 & 1.011& 3&$7^{+5.3}_{-4.2}\pm0.4$&4&$9^{+2.6}_{-2.1}\pm0.5$ & 4&$6^{+2.4}_{-1.9}\pm0.4$\\
		4.1574 & 407 & 5&$2^{+4.8}_{-4.0}$ & 11&$0^{+4.7}_{-3.9}$ & 0.175 & 0.318 &0.953& 6&$7^{+6.2}_{-5.2}\pm0.6$&7&$9^{+3.3}_{-2.8}\pm0.7$ & 7&$6^{+2.9}_{-2.5}\pm0.7$ \\
		4.1783 & 3189 & 24&$3^{+10.0}_{-9.3}$ & 46&$5^{+10.2}_{-9.9}$ & 0.174 & 0.329 & 0.918& 4&$2^{+1.7}_{-1.6}\pm0.4$&4&$3^{+0.9}_{-0.9}\pm0.4$ & 4&$3^{+0.8}_{-0.8}\pm0.4$ \\
		4.1888 & 566 & 2&$3^{+5.0}_{-4.3}$ & 7&$1^{+3.9}_{-3.2}$ & 0.176 & 0.327 & 0.942& 2&$2^{+4.7}_{-4.1}\pm0.2$&3&$6^{+2.0}_{-1.6}\pm0.3$ & 3&$4^{+1.8}_{-1.5}\pm0.3$  \\
		4.1989 & 526 & 9&$1^{+4.8}_{-4.2}$ & 4&$9^{+4.2}_{-3.4}$ & 0.170 & 0.318 & 1.012& 8&$9^{+4.7}_{-4.1}\pm0.8$&2&$6^{+2.2}_{-1.8}\pm0.2$ & 3&$6^{+2.0}_{-1.6}\pm0.3$ \\
		4.2092 & 517 & 5&$0^{+4.8}_{-4.0}$ & 4&$1^{+3.6}_{-2.9}$ & 0.155 & 0.294 & 1.134& 4&$8^{+4.7}_{-3.9}\pm0.5$&2&$1^{+1.9}_{-1.5}\pm0.2$ & 2&$5^{+1.7}_{-1.4}\pm0.2$ \\
		4.2187 & 515 & 6&$8^{+5.0}_{-4.3}$ & 8&$2^{+4.1}_{-3.4}$ & 0.142 & 0.270 & 1.290& 6&$4^{+4.7}_{-4.0}\pm0.6$&4&$0^{+2.0}_{-1.7}\pm0.4$ & 4&$4^{+1.8}_{-1.5}\pm0.4$  \\
		4.2263 & 1101 & 3&$8^{+5.8}_{-5.0}$ & 5&$5^{+4.7}_{-3.9}$ & 0.130 & 0.245 & 1.480& 1&$6^{+2.4}_{-2.1}\pm0.2$&1&$2^{+1.0}_{-0.9}\pm0.1$ & 1&$3^{+1.0}_{-0.8}\pm0.1$  \\
		4.2357 & 530 & -1&$3^{+3.3}_{-2.5}$ & -2&$2^{+2.0}_{-1.3}$ & 0.114 & 0.213 & 1.722& -1&$1^{+2.7}_{-2.1}\pm0.1$&-1&$0^{+0.9}_{-0.6}\pm0.1$ & -1&$0^{+0.9}_{-0.6}\pm0.1$  \\
		4.2438 & 538 & 5&$3^{+4.2}_{-3.5}$ & 0&$7^{+3.1}_{-2.3}$ & 0.096 & 0.183 & 1.999& 4&$5^{+3.6}_{-3.0}\pm0.4$&0&$3^{+1.4}_{-1.0}\pm0.1$ & 0&$8^{+1.3}_{-1.0}\pm0.1$  \\
		4.2580 & 828 & 3&$0^{+4.9}_{-4.1}$ & 5&$4^{+4.3}_{-3.5}$ & 0.076 & 0.141 & 2.610& 1&$6^{+2.6}_{-2.2}\pm0.2$&1&$6^{+1.2}_{-1.0}\pm0.1$ & 1&$6^{+1.1}_{-0.9}\pm0.2$ \\
		4.2668 & 531 & 6&$9^{+4.3}_{-3.6}$ & -3&$4^{+2.6}_{-2.0}$ & 0.062 & 0.121 & 3.075& 6&$0^{+3.8}_{-3.2}\pm0.6$&-1&$5^{+1.2}_{-0.9}\pm0.1$ & -0&$9^{+1.1}_{-0.9}\pm0.1$  \\
		4.2778 & 176 & \multicolumn{4}{c|}{$-4.2^{+3.4}_{-2.7}$} &\multicolumn{2}{c|}{ 0.074} & 3.602&\multicolumn{2}{c|}{-}&\multicolumn{2}{c|}{-}&-4&$0^{+3.2}_{-2.6}\pm0.3$  \\		
		4.2879 & 492 & -3&$7^{+3.0}_{-2.2}$ & -0&$4^{+3.0}_{-2.3}$ & 0.048 & 0.089 & 3.842& -3&$7^{+2.9}_{-2.1}\pm0.4$&-0&$2^{+1.6}_{-1.2}\pm0.1$ & -1&$0^{+1.4}_{-1.0}\pm0.1$ \\
		4.3079 & 45.1 & \multicolumn{4}{c|}{$1.0^{+2.0}_{-1.2}$} & \multicolumn{2}{c|}{0.086} & 2.708&\multicolumn{2}{c|}{-}&\multicolumn{2}{c|}{-}&4&$3^{+8.3}_{-5.2}\pm0.3$  \\		
		4.3121 & 492 & -0&$9^{+3.2}_{-2.5}$ & -1&$4^{+2.3}_{-1.6}$ & 0.069 & 0.125 & 2.407& -1&$0^{+3.5}_{-2.7}\pm0.1$&-0&$8^{+1.4}_{-1.0}\pm0.1$ & -0&$9^{+1.3}_{-0.9}\pm0.1$  \\
		4.3374 & 501 & 4&$2^{+4.4}_{-3.6}$ & 0&$5^{+2.4}_{-1.6}$ & 0.138 & 0.260 & 1.111& 4&$8^{+5.0}_{-4.2}\pm0.5$&0&$3^{+1.5}_{-1.0}\pm0.1$ & 0&$6^{+1.4}_{-0.9}\pm0.1$  \\
		4.3583 & 544 & 0&$8^{+3.9}_{-3.2}$ & 9&$0^{+4.0}_{-3.2}$ & 0.190 & 0.354 & 0.818& 0&$8^{+4.1}_{-3.3}\pm0.1$&5&$0^{+2.2}_{-1.8}\pm0.5$ & 4&$1^{+1.9}_{-1.6}\pm0.4$ \\
		
		4.3774 & 523 & 5&$5^{+4.0}_{-3.3}$ & 10&$4^{+4.3}_{-3.6}$ & 0.199 & 0.372 & 0.802& 5&$9^{+4.3}_{-3.5}\pm0.6$&5&$9^{+2.4}_{-2.0}\pm0.5$ & 5&$9^{+2.1}_{-1.8}\pm0.6$  \\
		4.3874 & 55.6 & \multicolumn{4}{c|}{$1.9^{+2.2}_{-1.5}$} & \multicolumn{2}{c|}{0.276} & 0.859&\multicolumn{2}{c|}{-}&\multicolumn{2}{c|}{-}&6&$3^{+7.4}_{-5.1}\pm0.4$  \\		
		4.3965 & 505 & 5&$9^{+4.0}_{-3.3}$ & 4&$2^{+3.2}_{-2.5}$ & 0.192 & 0.358 & 0.919& 5&$9^{+4.0}_{-3.3}\pm0.6$&2&$2^{+1.7}_{-1.3}\pm0.2$ & 2&$8^{+1.6}_{-1.2}\pm0.3$  \\
		4.4156 & 1091 & 9&$4^{+5.7}_{-4.9}$ & 14&$6^{+5.3}_{-4.6}$ & 0.171 & 0.324 & 1.066& 4&$2^{+2.5}_{-2.2}\pm0.4$&3&$4^{+1.2}_{-1.1}\pm0.3$ & 3&$6^{+1.1}_{-1.0}\pm0.3$ \\
		4.4362 & 568 & -5&$2^{+4.1}_{-3.4}$ & 6&$9^{+4.1}_{-3.3}$ & 0.153 & 0.288 & 1.192& -4&$3^{+3.4}_{-2.8}\pm0.4$&3&$0^{+1.8}_{-1.5}\pm0.3$ & 1&$4^{+1.6}_{-1.3}\pm0.1$ \\
		4.4671 & 111 & \multicolumn{4}{c|}{$-1.7^{+1.1}_{-0.7}$} & \multicolumn{2}{c|}{0.184} & 1.376&\multicolumn{2}{c|}{-}&\multicolumn{2}{c|}{-}&-2&$6^{+1.7}_{-1.1}\pm0.2$  \\		
		4.5271 & 112 &\multicolumn{4}{c|}{ $-0.2^{+2.1}_{-1.4}$ }& \multicolumn{2}{c|}{0.150} & 1.629&\multicolumn{2}{c|}{-}&\multicolumn{2}{c|}{-}&-0&$3^{+3.4}_{-2.2}\pm0.1$  \\
		
		4.5745 & 48.9 & \multicolumn{4}{c|}{$0.6^{+1.5}_{-0.8}$} & \multicolumn{2}{c|}{0.136} & 1.785& \multicolumn{2}{c|}{-}&\multicolumn{2}{c|}{-}&2&$1^{+5.5}_{-2.8}\pm0.1$  \\
		
		4.5995 & 587 & -1&$2^{+2.8}_{-2.0}$ & 1&$8^{+2.7}_{-1.9}$ & 0.093 & 0.173 &1.857& -1&$1^{+2.6}_{-1.8}\pm0.1$&0&$9^{+1.3}_{-0.9}\pm0.1$ & 0&$5^{+1.2}_{-0.8}\pm0.1$  \\

		\hline\hline
	\end{tabular}
\end{table*}

\begin{table*}[htbp]
	\centering
	\caption{Summary of the c.m. energy, luminosities, detection efficiencies, radiative correction factor, vacuum polarization factor, and the UL of born cross section ($90\%$ C.L.) of $\ee\to\gamma\chi_{c0}$.}
	\label{tab:chi0_chic0_upper}
	\begin{tabular}{c  |c |c|c |c |c|c|c}
		\hline\hline
		\centering{$\sqrt{s}$} & 	\centering{$\mathcal{L}_{int}(pb^{-1})$}&{\centering{$\epsilon_{K^+K^-\pi^+\pi^-}$}($\%$)} &{$\epsilon_{\pi^+\pi^-\pi^+\pi^-}$($\%$)}&\centering{$\epsilon_{K^+K^-}$($\%$)}& 1+$\delta$&1+$\delta_{v}$ &{\centering{$\sigma_{B}^{\rm up}$(pb)}}\\ 
		\hline
		4.0076&482 &22.7 &23.6&34.9&0.842&1.044&4.5  \\  
		4.1285&394 &20.2 &23.1&27.7&0.886&1.052&3.3 \\
		4.1574&407 &20.0 &22.9&27.4&0.892&1.053&7.7 \\	 			 		
		4.1783&3189&21.6 &22.9&33.7&0.897&1.054&2.2\\ 
		4.1888&566 &21.2 &22.7&33.1&0.899&1.056&4.3\\ 	
		4.1989&526 &21.4 &22.8&33.1&0.900&1.057&4.1\\
		4.2092&517 &21.4 &22.7&32.2&0.902&1.057&1.9\\	
		4.2187&515 &21.3 &22.8&32.8&0.904&1.056&4.8\\		
		4.2263&1101 &21.9 &23.0&32.6&0.905&1.056&2.5\\	
		4.2357&530&21.6&22.9&32.8&0.907&1.056&3.7\\	   	   	   	   	
		4.2438&538 &21.2 &22.8&32.8&0.908&1.055&2.4 \\	   	  
		4.2580&828 &21.4 &22.5&31.9&0.911&1.054&2.9\\	   	   	 	
		4.2668&531 &21.6 &23.1&33.2&0.912&1.053&1.8 \\	  
		4.2879&492 &19.6 &22.4&26.2&0.914&1.053&3.6 \\	
		
		4.3121&492 &19.7 &22.4&26.2&0.919&1.052&4.4 \\	
		4.3374&501 &19.7 &22.4&26.3&0.922&1.051&2.7\\
		
		4.3583&544 &21.4 &22.2&31.6&0.924&1.051&2.0 \\	 
		4.3774&523 &19.7 &22.3&26.4&0.926&1.051&2.6	\\		
		4.3965&505 &19.4 &22.5&25.8&0.928&1.051&5.4 \\			  	   	
		4.4156&1091 &21.4 &22.7&32.2&0.930&1.052&1.6\\	 
		4.4362&568 &19.4 &22.5&25.8&0.931&1.054&2.6 \\	   	   			  	   	
		4.5995&587 &20.5 &23.1&31.1&0.945&1.055&2.6\\	   	   	   	   	  	   			
		\hline\hline
	\end{tabular}
\end{table*}

\end{appendices}
\end{document}